\documentclass[journal]{IEEEtran}
\usepackage{cite}
\usepackage{amsmath,amssymb,amsfonts}
\usepackage{algorithm}  
\usepackage{algpseudocode}  
\usepackage{algpseudocode}
\usepackage{graphicx}
\usepackage{epstopdf}
\usepackage{textcomp}
\usepackage{xcolor}
\usepackage{bm}
\usepackage{setspace}
\ifCLASSINFOpdf
\else
\fi
\begin{document}
	
%
\title{Design and Optimization for Transmissive RIS Transceiver Enabled Uplink Communication Systems}
%
%
%
\author{Zhendong~Li,~Wen~Chen,~\IEEEmembership{Senior~Member,~IEEE},~Qingqing~Wu,~\IEEEmembership{Senior~Member,~IEEE},~Kunlun~Wang,~\IEEEmembership{Member,~IEEE},~and~Jun~Li,~\IEEEmembership{Senior~Member,~IEEE}
\thanks{Z. Li, W. Chen and Q. Wu are with the Department of Electronic Engineering, Shanghai Jiao Tong University, Shanghai 200240, China (e-mail: lizhendong@sjtu.edu.cn; wenchen@sjtu.edu.cn; qingqingwu@sjtu.edu.cn).}
\thanks{K. Wang is with the School of Communication and Electronic Engineering, East China Normal University, Shanghai 200241, China (e-mail: klwang@cee.ecnu.edu.cn).}
\thanks{J. Li is with the School of Electronic and Optical Engineering, Nanjing University of Science Technology, Nanjing 210094, China (email: jun.li@njust.edu.cn). }
\thanks{(\emph{Corresponding author: Wen Chen.})}
}
\maketitle

\begin{abstract}
	In this paper, a novel transmissive reconfigurable intelligent surface (RIS) enabled uplink communication system with orthogonal frequency division multiple access (OFDMA) is investigated. Specifically, a non-conventional receiver architecture equipped with a single receiving horn antenna and a transmissive RIS is first proposed, and a far-near field channel model based on planar waves and spherical waves is also given. Then, in order to maximize the system sum-rate of uplink communications, we formulate a joint optimization problem over subcarrier allocation, power allocation and RIS transmissive coefficient design while taking account of user quality-of-service (QoS) constraint. Due to the coupling of optimization variables, the optimization problem is non-convex, so it is challenging to solve it directly. In order to tackle this problem, the alternating optimization (AO) algorithm is utilized to decouple the optimization variables and divide the problem into two sub-problems to solve. First, the problem of joint subcarrier allocation and power allocation is solved via the Lagrangian dual decomposition method. Then, the RIS transmissive coefficient design scheme can be obtained by applying difference-of-convex (DC) programming, successive convex approximation (SCA) and penalty function methods. Finally, the two sub-problems are iterated alternately until convergence is achieved. Numerical results verify that the proposed algorithm has good convergence performance and can improve sum-rate of the proposed system compared with other benchmark algorithms.
\end{abstract}


%
\IEEEpeerreviewmaketitle

\section{Introduction}
%
%
%
%
\IEEEPARstart{W}{ith} the vigorous development and continuous evolution of wireless communication technology, the power consumption and cost of base station (BS) are constantly increasing. Taking the fifth-generation (5G) networks for instance, the power consumption of a single 5G BS is about 3500W, while the power consumption of a single fourth-generation (4G) BS is about 1300W. Thus, the single station power consumption of 5G is about three times that of 4G. Meanwhile, due to the higher frequency of 5G networks, to achieve the same coverage as 4G networks, the number of 5G BSs is about three to four times that of 4G BSs, so the deployment costs are quite high. In addition, since the massive multiple-input multiple-output (MIMO) system of 5G networks require a large number of radio frequency (RF) chains and complex signal processing units, the cost of a single station is also greatly increased \cite{7894280,9113273}. Therefore, it is urgent to seek a novel multi-antenna system that can reduce power consumption and cost for future communication networks.

Recently, as one of the innovative technologies, reconfigurable intelligent surface (RIS) has attracted attention from academia and industry \cite{8811733,9167258,8741198}. RIS composed of a large number of passive elements can reconstruct wireless channels by changing the amplitude and phase of incident electromagnetic (EM) waves. RIS can reduce the strength of the interference signal and increase the strength of the desired signal when it used to assist communication, thereby enhancing the performance of the communication system. Due to the almost passive full-duplex operation mode, the RIS only passively reflects or transmits the incident EM waves, so compared with the traditional relay technology, RIS does not encounter the problem of self-interference and additional noise \cite{9167258}. Since it does not contain complicated signal processing modules, so the required hardware cost and power consumption are greatly reduced compared to the traditional multi-antenna system \cite{9117136}. These advantages make RIS have great potential in future wireless networks.

In the current research, RIS mainly has two modes: reflective mode \cite{9531372,9887822,9509394,9913311,9174801,9110869,9226616} and transmissive mode \cite{9365009,9855406,zhang2021secrecy,aldababsa2021simultaneous,zuo2021uplink}. At present, more research focuses on the RIS-assisted communication in reflective mode or transmissive mode. The RIS in reflective mode is mainly used to improve the performance of the networks. Li \emph{et al.} investigated intelligent reflecting surface (IRS)-assisted simultaneous wireless information and power transfer (SWIPT) non-orthogonal multiple access (NOMA) system, where IRS is applied to improve the NOMA performance and the wireless power transfer (WPT) efficiency of SWIPT \cite{9531372}. Chen \emph{et al.} considered IRS-aided wireless-powered mobile edge computing (WP-MEC) system, where time division multiple access (TDMA) and non-orthogonal multiple access (NOMA) are both considered in uplink transmission, and performance improvement is achieved through IRS \cite{9887822}. Liu \emph{et al.} formulated a problem of joint deployment, IRS phase shift and power allocation in the multiple-input single-output (MISO) NOMA networks to maximize the energy efficiency subject to the user's data requirements \cite{9174801}. Huang \emph{et al.} proposed a joint design of BS transmit beamforming and reconfigurable intelligent surface (RIS) phase shift matrix by applying deep reinforcement learning (DRL) \cite{9110869}. Wang \emph{et al.} used IRS to provide effective reflective paths to enhance the coverage of mmWave networks, where the user receiving signal power is maximized by jointly optimizing BS transmit beamforming and IRS phase shift \cite{9226616}.

Moreover, the transmissive RIS is mainly used to solve the blind coverage problem and improve the network coverage. Zeng \emph{et al.} considered a downlink multi-user RIS-assisted communication network, where the RIS can be chosen three modes. The system sum-rate is derived, and which mode can bring the best performance with a specific user distribution also be discussed \cite{9365009}. Li \emph{et al.} studied coverage probability of the multiple transmissive RIS-aided future outdoor-to-indoor (O2I) millimeter wave (mmWave) networks, where the transmissive RIS is used to enhance coverage \cite{9855406}. Aldababsa \emph{et al.} investigated simultaneously transmitting and reflecting reconfigurable intelligent surfaces (STAR-RIS) assisted NOMA communication system, where the STAR-RIS utilizes the mode switching (MS) protocol to improve coverage and serve all users located on both sides of the RIS \cite{aldababsa2021simultaneous}. The above-mentioned researches are all based on the perspective of RIS with different modes for communication assistance.

In addition to being used for auxiliary communication, RIS can also be considered as transceiver for communication. These two research perspectives are completely different. The latter is in its infancy and also very promising. Currently, some studies have proposed that the reflective RIS can be used as a transmitter \cite{9133266}. However, unlike the reflective RIS transmitter structure, the receiving horn antenna and the user are located on opposite sides of the RIS under the transmissive RIS transceiver architecture, so the receiving horn antenna will not cause the problems of feed source occlusion and echo interference. Therefore, transmissive RIS transceivers can be designed more efficiently \cite{bai2020high,wan2021space,liu2021multifunctional,7448838}, which is very promising in future communication networks. Accordingly, we have proposed a transmissive RIS transmitter architecture in \cite{9570775}, and conducted a preliminary study on the downlink multi-user beamforming design (i.e., RIS transmissive coefficient design) and power allocation under the transmissive RIS transmitter architecture. Our goal is to propose a novel transmissive RIS transceiver architecture (i.e., the design of downlink communication and uplink communication), which can greatly reduce power consumption and cost. In addition, the uplink and downlink communication mechanisms of the transmissive RIS transceiver are completely different, and there is also no clear solution for the uplink communication design of the reflective RIS transmitter. Therefore, these have greatly facilitated our research on the transmissive RIS transceiver enabled uplink communication system in this paper.

Since the proposed transmissive RIS transceiver is equipped with a single receiving horn antenna, we consider orthogonal frequency division multiple access (OFDMA) for uplink multi-user communications. For a multi-carrier system, the subcarrier allocation of this access scheme is more flexible, and it can bring higher frequency diversity gain \cite{7543459,8809094}. There are many works to optimize the power consumption, spectrum- and energy- efficiency of the network through joint design of OFDMA network configuration \cite{7208841,5582313,7515184,8438896,7062017,5783982,7154414,7138584}. For uplink communication of a single-cell OFDMA system, Souza \emph{et al.} solved energy efficiency maximization by joint optimizing power allocation and subcarrier allocation for all mobile users \cite{7543459}. For uplink OFDMA based cloud radio access network (C-RAN), Liu \emph{et al.} jointly optimized power allocation and fronthaul quantization design on the subcarrier to maximize the system throughput with practical uniform scalar quantization \cite{7208841}. Chen \emph{et al.} studied the downlink throughput maximization for OFDMA networks with feedback channel capacity constraints \cite{5582313}. Wang \emph{et al.} proposed resource allocation algorithm for multi-cell OFDMA system with imperfect channel state information (CSI) \cite{7062017}. Xiong \emph{et al.} investigated energy-efficient resource allocation algorithm for OFDMA two-way relay \cite{7154414}. Li \emph{et al.} solved a max-min energy efficiency resource allocation problem in OFDMA systems, where the energy efficiency of the worst-case link is maximized by jointly optimizing transmit power, user's QoS and subcarrier allocation \cite{7138584}.

In this paper, we mainly study the design and optimization issues of our proposed transmissive RIS transceiver enabled uplink communication. Specifically, we aim to elaborate the proposed architecture and solve a problem of maximizing system sum-rate by jointly optimizing multi-user power allocation, subcarrier allocation and RIS transmissive coefficient. Due to the high coupling of optimization variables, this problem is a non-convex optimization problem, so it is difficult to obtain the optimal solution. Therefore, we need to design an effective algorithm to obtain the suboptimal solution and verify it by numerical simulations. In summary, the main contributions of this paper are as follows:
\begin{itemize}
\item We first propose a transmissive RIS transceiver enabled uplink communication system, where the new transceiver is composed of receiving horn antenna and transmissive RIS. Since the receiving horn antenna is a single antenna, when the transceiver provides uplink services for multiple users, the users can use OFDMA to access. Then, through the design of the RIS transmissive coefficient, the uplink transmission signals of users can be strengthened. Specially, we establish an optimization problem that maximizes system sum-rate by jointly optimizing multi-user power allocation, subcarrier allocation and RIS transmissive coefficient while taking into account the QoS requirement of each user. Since the optimization variables are coupled, the problem is non-convex. Accordingly, it is challenging to obtain the optimal solution directly.

\item In order to address the system sum-rate maximization problem, we use an alternating optimization (AO) algorithm framework to divide it into two sub-problems. In the first sub-problem, given the RIS transmissive coefficient, the multi-user power allocation and subcarrier allocation scheme can be jointly obtained by applying Lagrangian dual decomposition method. In the second sub-problem, based on the obtained power allocation and subcarrier allocation scheme, through difference-of-convex (DC) programming and penalty function method, the RIS transmissive coefficient design problem can be transformed into a semi-definite programming (SDP) problem to solve. Finally, the two sub-problems are iterated alternately until convergence is achieved.

\item Through numerical simulation, we verify the effectiveness of the proposed system sum-rate maximization algorithm of joint multi-user power allocation, subcarrier allocation and RIS transmissive coefficient compared with the baseline algorithm, i.e., it can improve the sum-rate of the system. Meanwhile, the number of transmissive RIS elements also has an impact on system performance. When the number of RIS elements is larger, the sum-rate of the system is higher, which has great potential for future ultra massive MIMO in 6G networks.
\end{itemize}

The remainder of this paper is organized as follows. Section II elaborates the the advantages and transmission mechanism of transmissive RIS transceiver in detail. Section III introduces the system model, far-near field channel model and optimization problem formulation for the uplink transmissive RIS transceiver enabled communication system with OFDMA. Section IV further elaborates the proposed joint subcarrier allocation, power allocation and RIS transmissive coefficient design algorithm for the formulated non-convex optimization problem. The computational complexity analysis and convergence analysis of the algorithm are also given. Then, in Section V, numerical results demonstrate that proposed algorithm has good convergence and effectiveness. Finally, the conclusion is given in Section VI.

\textit{Notations:} In this paper, scalars, vectors and matrices are respectively represented by lower-case letters, bold lower-case letters and bold upper-case letters. The absolute value of a complex-valued scalar $x$ can be denoted by $\left| {x} \right|$, and the Euclidean norm of a complex-valued vector $\bf{x}$ can be denoted by $\left\| {\bf{x}} \right\|$. In addition, $\rm{tr(\bf{X})}$, $\rm{rank(\bf{X})}$, ${\bf{X}}^H$, ${\bf{X}}_{m,n}$ and $\left\| {\bf{X}} \right\|$ denote trace, rank, conjugate transpose, $m,n$-th entry and matrix norm of a square matrix $\bf{X}$, respectively, while ${\bf{X}} \succeq 0$ represents the square matrix $\bf{X}$ is a positive semidefinite matrix. Similarly, $\rm{rank(\bf{A})}$, ${\bf{A}}^H$, ${\bf{A}}_{m,n}$ and $\left\| {\bf{A}} \right\|$ also denote rank, conjugate transpose, $m,n$-th entry and matrix norm of a general matrix $\bf{A}$, respectively. ${\bf{X}} \otimes {\bf{Y}}$ represents the Kronecker product of two matrices ${\bf{X}}$ and ${\bf{Y}}$. ${\mathbb{C}^{M \times N}}$ represents the space of ${M \times N}$ complex matrix. $j$ represents the imaginary unit, i.e., $j^2=-1$. Finally, the distribution of a circularly symmetric complex Gaussian (CSCG) random vector with mean $\mu$ and covariance matrix $\bf{C}$ can be expressed as $ {\cal C}{\cal N}\left( {\mu,\bf{C}} \right)$, and $\sim$ denotes `distributed as'.
\section{The Advantages and Transmission Mechanism of Transmissive RIS Transceiver Enabled System}
\subsection{The Advantages of Transmissive RIS Transceiver}
In this subsection, we elaborate on the advantages of our proposed transmissive RIS transceiver compared to the other multi-antenna system. Compared with traditional multi-antenna systems, this architecture does not require a large number of radio frequency (RF) chains, complex signal processing modules, etc. Hence, the new transceiver architecture is low-power consumption and low-cost. Then, compared with the reflective RIS transmitter, it mainly has the following two advantages.

{\bf{(1) No feed occlusion:}} The user and horn antenna of the reflective RIS transmitter system are located on the same side of the RIS, while the user and horn antenna of the transmissive RIS transceiver system are located on opposite sides of the RIS. Therefore, there is no feed occlusion problem in the transmissive RIS transceiver system.

{\bf{(2) No self-interference problem:}} For the reflective RIS transmitter system, the incident EM waves and the reflective EM waves are on the same side of the RIS, which will bring self-interference problems. As for the transmissive RIS transceiver system, the incident EM waves and the transmissive EM waves are on the opposite sides of the RIS, which will not cause self-interference problems. Therefore, the transmissive RIS transceiver system is more promising for complex communication environments in the future.
\begin{figure}
	\centerline{\includegraphics[width=9cm]{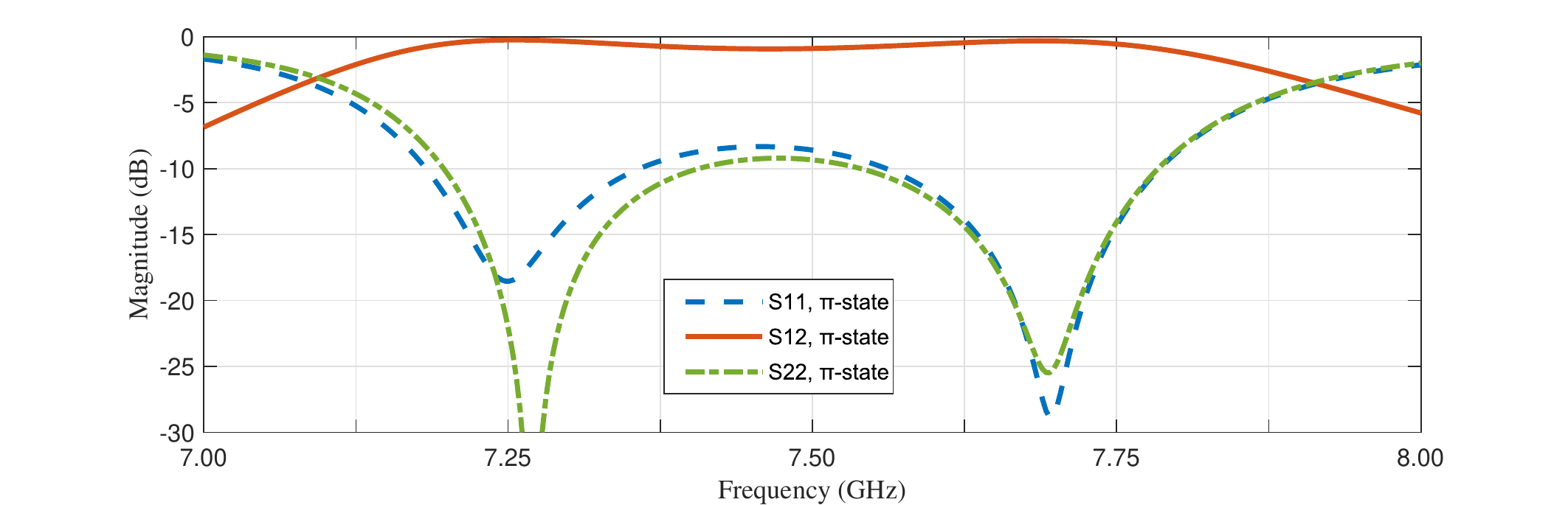}}
	\caption{Magnitude for $\pi$-state.}
	\label{Fig1}
\end{figure}
\begin{figure}
	\centerline{\includegraphics[width=9cm]{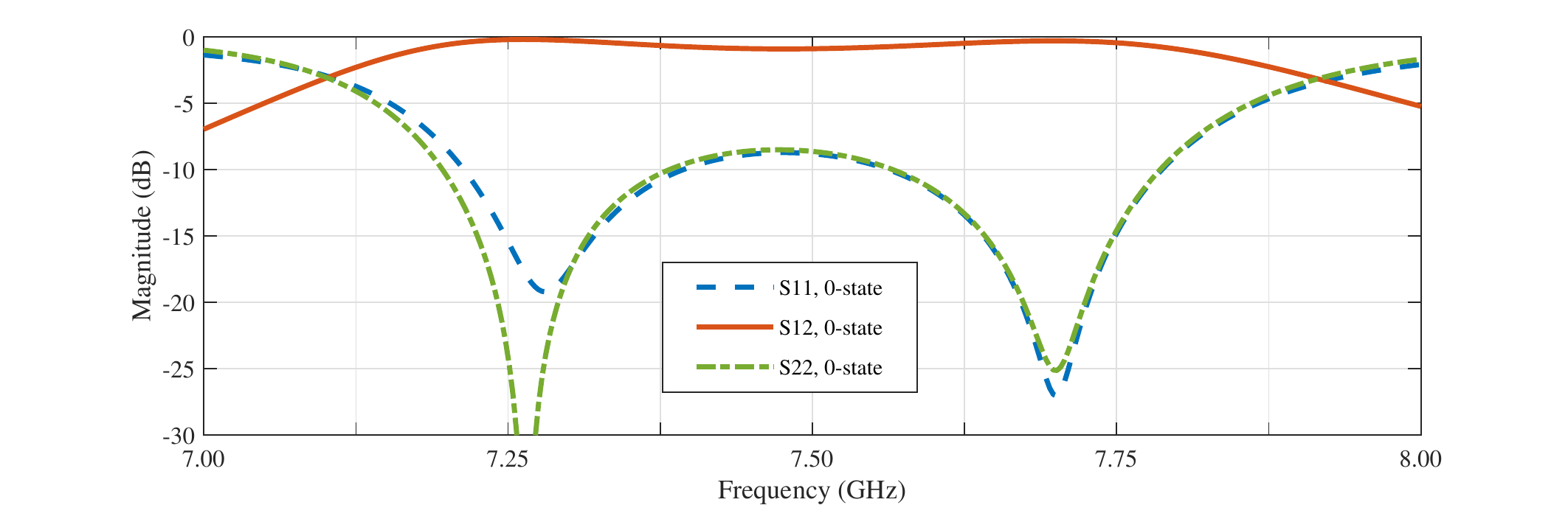}}
	\caption{Magnitude for 0-state.}
	\label{Fig1}
\end{figure}

Based on the advantages brought by the above physical structure, the transmissive RIS transceiver can achieve higher aperture efficiency and operating bandwidth \cite{bai2020high}. According to literature \cite{bai2020high}, Fig. 1 and Fig. 2 are the magnitudes for the simulated scattering coefficients of 1-bit transmitted RIS under different phase states, where $S_{11}$ and $S_{22}$ represent the case of reflective EM. Moreover, $S_{12}$ represents the case of transmissive EM. It can be seen that from 7.12 to 7.85 GHz, in both 0-state and $\pi$-state, the transmissive coefficient is above -2dB, and the reflective coefficient is below -10dB. Thus, higher aperture efficiency and operating bandwidth can be achieved with lower power consumption and cost under this transmissive RIS transceiver architecture.

\subsection{Transmission Mechanism of Transmissive RIS Transceiver}
{\bf{(1) Downlink transmission:}} We have proposed the transmissive RIS transmitter enabled downlink transmission process in \cite{9570775}. Specifically, the any-order modulation information from the signal source can be modulated in the controller by time-modulation array (TMA) \cite{9133266}, and it is used as one part of the RIS transmissive coefficient. Then, the beamforming design proposed by \cite{9570775} is used as the other part of the RIS transmissive coefficient. The mixed signal is the RIS transmissive coefficient. In other words, the RIS controller can realize information modulation and beamforming design through the corresponding control signal, and the coefficient can be mapped to each element of the transmissive RIS. Then, the horn antenna sends a single-frequency EM wave to carry and send the information of each element to multiple users, and the multiuser diversity can be achieved by using space division multiple access (SDMA).

{\bf{(2) Uplink transmission:}} In the uplink architecture proposed in this paper, since the receiving horn antenna is equipped with a single antenna, we consider multi-user access by using OFDMA. First, on each subchannel, the user sends modulation signal, and the transmissive RIS strengthens the uplink signal and forwards it to the receiving horn antenna. Fianlly, the receiving horn antenna sends the received signal to the controller for demodulation and decoding. This paper is devoted to the performance improvement of the transmissive RIS transceiver enabled uplink communication system. Next, we will focus on how to design and optimize this uplink transmission architecture.
\section{System Model and Problem Formulation}
\subsection{System Model}
As shown in Fig. 3, we consider an uplink multi-user OFDMA network based on the transmissive RIS transceiver. The network consists of a receiving horn antenna, RIS with ${M_c} \times {M_r}$ transmissive elements, and $K$ single-antenna users. Let ${\bf{c}} = {\left[ {{c_{1,1}},...,{c_{{M_c},{M_r}}}} \right]^T} \in {\mathbb{C}^{{M_c}{M_r} \times 1}}$ denote the RIS transmissive coefficient vector, where ${c_{{m_c},{m_r}}} = {\beta _{{m_c},{m_r}}}{e^{j{\theta _{{m_c},{m_r}}}}}$. ${\beta _{{m_c},{m_r}}} \in \left[ {0,1} \right]$ and ${\theta _{{m_c},{m_r}}} \in \left[ {0,2\pi } \right)$ respectively represent the transmissive amplitude and phase shift of the $(m_c,m_r)$-th element of RIS. The RIS transmissive coefficient needs to meet
\begin{equation}
	\left| {{c_{{m_c},{m_r}}}} \right| \le 1,\forall m_c,m_r.
\end{equation}
The controller equipped with RIS can adjust the amplitude and phase shift of the transmissive element to achieve the enhancement of the uplink incident signal. To facilitate the analysis, we assume that the RIS is fully transmissive, i.e., no incident signal is reflected. In this paper, the elements of RIS are modeled in the way of uniform planar array (UPA). Specifically, each column has $M_c$ elements arranged at equal spacing $d_c$ meters, and each row has $M_r$ elements arranged at equal spacing $d_r$ meters.
\begin{figure}
	\centerline{\includegraphics[width=9cm]{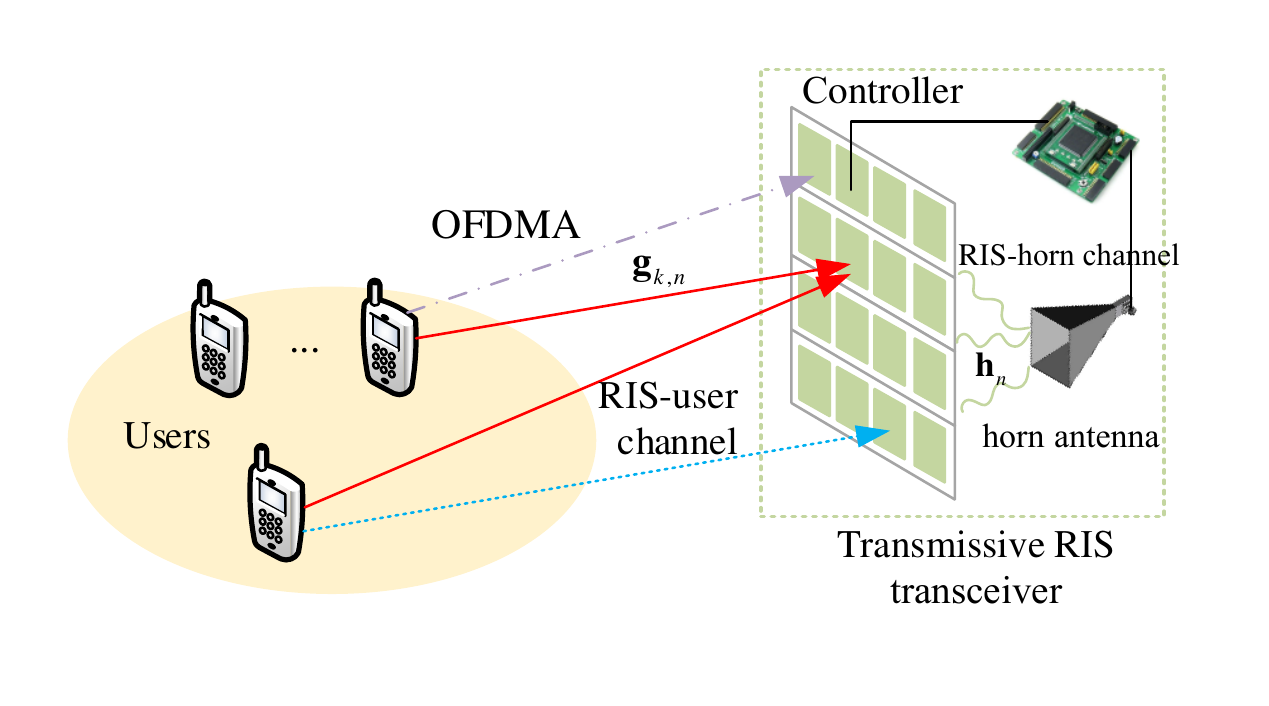}}
	\caption{Transmissive RIS transceiver enabled uplink multi-user communication systems with OFDMA.}
	\label{Fig3}
\end{figure}

In this paper, we consider that the channel with bandwidth $B$ is divided into $N$ subcarriers, and the bandwidth of each subcarrier is $W = {B \mathord{\left/
		{\vphantom {B N}} \right.
		\kern-\nulldelimiterspace} N}$. Inter-subcarrier interference can be ignored, and the cyclic prefix is large enough to overcome inter-symbol interference. Next, we describe the broadband far-near field channel model of the proposed system. As shown in Fig. 4, the EM wave radiation field in the wireless communication network can be divided into far field and near field \cite{cui2021channel}, which have different channel models. The far-near field is determined by the Rayleigh distance $\frac{{2{D^2}}}{\lambda }$ \cite{7942128}, where $D$ and $\lambda$ respectively denote array aperture and wavelength. The judgment method by applying Rayleigh distance is that, if the distance between the transmitter and the receiver is larger than Rayleigh distance, the channel can be regarded as a far-field channel, and the wavefront can be approximated as a planar wave. Otherwise, when the distance between the the transmitter and the receiver is less than Rayleigh distance, the channel can be modeled as near-field channel, and the wavefront is spherical wave.
\begin{figure}
	\centerline{\includegraphics[width=10cm]{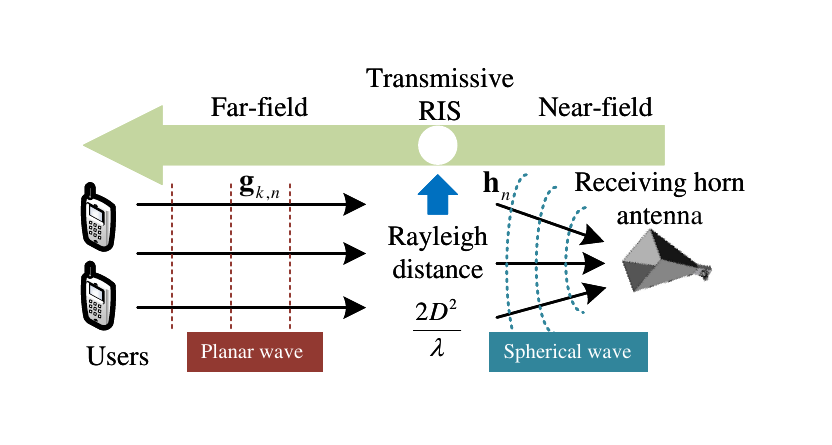}}
	\caption{Far-near field region separated by Rayleigh distance \cite{cui2021channel}.}
	\label{Fig4}
\end{figure}

{\bf{(1) Far-field RIS-user channel model:}} On the $n$-th subcarrier, the channel from the $k$-th user to the RIS is denoted as ${{\bf{g}}_{k,n}} \in {\mathbb{C}^{M_cM_r \times 1}},\forall k,n$, which can be named RIS-user channel. Since the distance from the user to the RIS is larger than the Rayleigh distance, the RIS-user channel can be regarded as a far-field channel, which is modeled under the assumption of planar waves. Considering that there are both line-of-sight (LoS) components and non-line-of-sight (NLoS) components between the user and the RIS, we model the RIS-user channel as a Rician fading channel. Hence, the channel gain from the $k$-th user to the RIS on the $n$-th subcarrier can be expressed as
\begin{equation}
	\begin{aligned}
		{{\bf{g}}_{k,n}} \!= \!\sqrt {\frac{{{C_0}}}{{{{\left( {{d_k}} \right)}^\alpha }}}} \left( {\sqrt {\frac{\kappa }{{1 \!+\! \kappa }}} {e^{ - j2\pi nW\frac{{{d_k}}}{c}}}{\bf{g}}_k^{{\rm{LoS}}} \!+\! \sqrt {\frac{1}{{1\! +\! \kappa }}} {\bf{g}}_{k,n}^{{\rm{NLoS}}}} \right),\\\forall k,n,
	\end{aligned}
\end{equation}
with ${\bf{g}}_{k,n}^{{\rm{NLoS}}} \sim {\cal C}{\cal N}\left( {{\bf{0}},{{\bf{I}}_{{M_c}{M_r}}}} \right)$ and ${{\bf{g}}_{k}^{{\rm{LoS}}}}$ can be given by
\begin{equation}
	\begin{aligned}
		{\bf{g}}_{k}^{{\rm{LoS}}} \!=\! &{\left[ {1,{e^{ - j2\pi {f_c}\frac{{{d_r}\sin {\theta _k}\cos {\psi _k}}}{c}}},\!...,{e^{ - j2\pi {f_c}\left( {{M_r}\! -\! 1} \right)\frac{{{d_r}\sin {\theta _k}\cos {\psi _k}}}{c}}}} \right]^T} \\
		\otimes& {\left[ {1,{e^{ - j2\pi {f_c}\frac{{{d_c}\sin {\theta _k}\sin {\psi _k}}}{c}}},\!...,{e^{ - j2\pi {f_c}\left( {{M_c} \!- \!1} \right)\frac{{{d_c}\sin {\theta _k}\sin {\psi _k}}}{c}}}} \right]^T},
	\end{aligned}
\end{equation} 
where $C_0$ is channel gain when reference distance is 1 meter, $d_k$ denotes the distance between RIS and the $k$-th user, $\alpha$ denotes path loss exponent of the RIS-user channel, $c$ represents speed of light, $\kappa$ is Rician factor and $f_c$ denotes carrier frequency. ${\theta _k}$ and ${{\psi _k}}$ respectively denote vertical and horizontal angle-of-arrival (AoA). It is worth noting that usually $B \ll  {f_c}$ holds, so Eq. (3) only depends on the corresponding AoA, so it is frequency flat, i.e., it is not determined by the index of the subcarrier. However, the phase shift term ${{e^{ - j2\pi nW\frac{{{d_k}}}{c}}}}$ depends on the index of the subcarriers, even when they share the same delay ${\frac{{{d_k}}}{c}}$\footnote{The frequency domain representation of the discrete LoS channel $\delta \left[ {m - {m_\tau }} \right]$ with a delay of $\tau  = \frac{{{m_\tau }}}{B}$ can be obtained by performing $N$-point Discrete Fourier Transform (DFT), which can be given by ${\rm{DFT}}\left\{ {\delta \left[ {m - {m_\tau }} \right]} \right\} = {e^{ - j\frac{{2\pi n}}{N}{m_\tau }}} = {e^{ - j2\pi nW\tau }},\forall n = 0,...,N - 1$, where $\delta \left[  \cdot  \right]$ denotes delta function.}. Accordingly, all $N$ subcarriers have different phase \cite{9293155}.

{\bf{(2) Near-field RIS-receiving channel model:}} On the $n$-th subcarrier, the channel from the RIS to the receiving antenna can be denoted as ${{\bf{h}}_n} \in {\mathbb{C}^{{M_c}{M_r} \times 1}},\forall n$, which can be named RIS-receiving channel. Since the distance from the receiving antenna to the RIS is less than the Rayleigh distance, the RIS-receiving channel can be regarded as a near-field channel, which is modeled under the assumption of spherical waves. Considering that there is no occlusion between the RIS and the receiving antenna, we model it as a LoS channel, which can be expressed as
\begin{equation}
	{{\bf{h}}_n} = \rho {e^{ - j2\pi nW\frac{\tilde r}{c}}}{{\bf{h}}^{{\rm{LoS}}}},\forall n,
\end{equation}
with ${{\bf{h}}^{{\rm{LoS}}}}$ can be given by
\begin{equation}
	\begin{aligned}
		{{\bf{h}}^{{\rm{LoS}}}} = \left[ {{e^{ - j2\pi {f_c}\frac{{{r_{1,1}} - \tilde r}}{c}}},...{e^{ - j2\pi {f_c}\frac{{{r_{{m_c},{m_r}}} - \tilde r}}{c}}}},\right.\\
				\left.{...,{e^{ - j2\pi {f_c}\frac{{{r_{{M_c},{M_r}}} - \tilde r}}{c}}}} \right]^H,
	\end{aligned}
\end{equation}
where $\rho$ and $\tilde r$ denote complex gain and the the distance from the center of the RIS to the receiving horn antenna, respectively. ${{r_{{m_c},{m_r}}}}$ is the distance from the (${{m_c},{m_r}}$)-th RIS element to the receiving antenna, which can be given by
\begin{equation}
	{r_{{m_c},{m_r}}} = \sqrt {{{\tilde r}^2} + \hat d_{{m_c},{m_r}}^2} ,
\end{equation}
where ${{\hat d}_{{m_c},{m_r}}} = \sqrt {\delta _{{m_c}}^2d_c^2 + \delta _{{m_r}}^2d_r^2} $ denotes the distance from the (${{m_c},{m_r}}$)-th RIS element to the center of the RIS, ${\delta _{{m_c}}} = \frac{{2{m_c} - {M_c} - 1}}{2}$ and ${\delta _{{m_r}}} = \frac{{2{m_r} - {M_r} - 1}}{2}$.

On the $n$-th subcarrier, the signal of the $k$-th user received by receiving horn antenna can be expressed as
\begin{equation}
	{y_{k,n}} = \sqrt {{p_{k,n}}} {{\bf{h}}_n^H}{\rm{diag}}\left( {{{\bf{g}}_{k,n}}} \right){\bf{c}}{x_{k,n}} + n_0,\forall k,n,
\end{equation}
where ${x_{k,n}}$ represents the transmitting signal of the $k$-th user on the $n$-th subcarrier, which is assumed to be an independent and identically distributed CSCG random variable with with zero mean and unit variance, i.e., ${x_{k,n}} \sim {\cal C}{\cal N}\left( {0,1} \right)$. $n_0$ denotes the additive white Gaussian noise (AWGN) introduced by the receiving horn antenna, i.e., $n_0 \sim {\cal C}{\cal N}\left( {0,{\sigma ^2}} \right)$. Herein, ${\sigma ^2} = {N_0}W$, $N_0$ is the noise power spectral density (PSD). Therefore, on the $n$-th subcarrier, the signal-to-noise ratio (SNR) of the $k$-th user is
\begin{equation}
	{\gamma _{k,n}} = \frac{{{p_{k,n}}{{\left| {{{\bf{h}}_n^H}{\rm{diag}}\left( {{{\bf{g}}_{k,n}}} \right){\bf{c}}} \right|}^2}{\nu _k}}}{{{N_0}W}},\forall k,n,
\end{equation}
where ${p_{k,n}}$ is the transmit power of the $k$-th user on the $n$-th subcarrier, and ${\nu _k}$ describes the imperfections of the channel and modulation \cite{7543459}, which can be expressed as
\begin{equation}
	{\nu _k} = \frac{{ - 1.5}}{{\log \left( {5{\rm{BE}}{{\rm{R}}_k}} \right)}},\forall k,
\end{equation}
where ${\rm{BE}}{{\rm{R}}_k}$ denotes the maximum allowable bit error rate of the $k$-th user.

According to Shannon’s formula, the bit rate of the $k$-th user on the $n$-th subcarrier can be expressed as
\begin{equation}
	{r_{k,n}} = W{\log _2}\left( {1 + {\gamma _{k,n}}} \right),\forall k,n.
\end{equation}
The total bit rate of the system can be further expressed as
\begin{equation}
	R\left( {{\bf{A}},{\bf{P}},{\bf{c}}} \right) = W\sum\limits_{k = 1}^K {\sum\limits_{n = 1}^N {{a_{k,n}}{{\log }_2}\left( {1 + {\gamma _{k,n}}} \right)} } ,
\end{equation}
where ${\bf{A}} = \left[ {{a_{k,n}}} \right],\forall k,n$ is subcarrier allocation matrix, ${a_{k,n}} \in \left\{ {0,1} \right\}$. When the $n$-th subcarrier is allocated to the $k$-th user, ${a_{k,n}} = 1$. Otherwise, ${a_{k,n}} = 0$. It should satisfy the following constraints
\begin{equation}
	\sum\limits_{k = 1}^K {{a_{k,n}} \le 1,} \forall n.
\end{equation}
In addition, ${\bf{P}} = \left[ {{p_{k,n}}} \right],\forall k,n$ denotes power allocation matrix, which should meet the following constraints
\begin{equation}
	{p_{k,n}} \ge 0,\forall k,n,
\end{equation}
and
\begin{equation}
	\sum\limits_{n = 1}^N {{a_{k,n}}{p_{k,n}}}  \le p_k^{\max },\forall k.
\end{equation}
\newcounter{my3}
\begin{figure*}[!t]
	\normalsize
	\setcounter{my3}{\value{equation}}
	\setcounter{equation}{17}
	\begin{equation}
		\begin{aligned}
			{\cal L}\left( {{\bf{\tilde P}},{\bf{\tilde A}},{\bm{\lambda }},{\bm{\mu }}} \right) =& W\sum\limits_{k = 1}^K {\sum\limits_{n = 1}^N {{{\tilde a}_{k,n}}{{\log }_2}\left( {1 + \frac{{{{\tilde p}_{k,n}}{\Gamma _{k,n}}}}{{{{\tilde a}_{k,n}}}}} \right)} }  - \sum\limits_{k = 1}^K {{\lambda _k}\left( {\sum\limits_{n = 1}^N {{{\tilde p}_{k,n}}}  - p_k^{\max }} \right)} \\
			- &\sum\limits_{k = 1}^K {{\mu _k}\left( {r_k^{\min } - \sum\limits_{n = 1}^N {{{\tilde a}_{k,n}}{{\log }_2}\left( {1 +\frac{{{{\tilde p}_{k,n}}{\Gamma _{k,n}}}}{{{{\tilde a}_{k,n}}}}} \right)} } \right)} ,
		\end{aligned}
	\end{equation}
	\setcounter{equation}{\value{my3}}
	\hrulefill
	\vspace*{4pt}
\end{figure*}
\subsection{Problem Formulation}
In this paper, our goal is to maximize the sum-rate of the proposed system by jointly optimizing the subcarrier allocation $\bf{A}$, power allocation $\bf{P}$ and RIS transmissive coefficient $\bf{c}$ in the transmissive RIS transceiver enabled uplink communication system with OFDMA. Therefore, the optimization problem can be formulated as follows,
\begin{subequations}
	\begin{align}
		\left( {{\textrm{P0}}} \right){\rm{~~~~~}}&\mathop {\max}\limits_{{\bf{A}},{\bf{P}},{\bf{c}}} {\rm{ ~~~}}R\left( {{\bf{A}},{\bf{P}},{\bf{c}}} \right), \nonumber\\ 
		\rm{s.t.}\qquad &{p_{k,n}} \ge 0,\forall k,n,\\
		&{\rm{ }}\sum\limits_{n = 1}^N {{a_{k,n}}{p_{k,n}}}  \le p_k^{\max },\forall k,\\
		&{a_{k,n}} \in \left\{ {0,1} \right\},\forall k,n,\\
		&\sum\limits_{k = 1}^K {{a_{k,n}} \le 1,} \forall n,\\
		&\sum\limits_{n = 1}^N {{a_{k,n}}{r_{k,n}}}  \ge r_k^{\min },\forall k,\\
		&\left| {{c_m}} \right| \le 1,\forall m,
	\end{align}
\end{subequations}
where (15a) means that all users must have positive power allocated on any subcarrier, (15b) means that the transmit power of the $k$-th user cannot exceed its maximum transmit power, and (15c) represents that each subcarrier allocation has only two states. (15d) ensures that each subcarrier is allocated to at most one user, (15e) is the QoS constraint for each user, and (15f) represents the constraint of the RIS transmissive coefficient.

The problem (P0) is a non-convex optimization problem, mainly due to the following reasons: First, the three optimization variables are highly coupled, so the objective function is non-concave. In addition, (15c) is an integer constraint, which is a non-convex constraint, so (15b)-(15e) are all non-convex constraints. Therefore, the problem (P0) is a non-convex optimization problem, and it is not easy to solve it directly. Therefore, it is necessary to design an algorithm to solve the sum-rate maximization problem. The algorithm for solving the problem (P0) will be introduced in next section.

From the optimization problem (P0), we can see that the uplink RIS transmissive coefficient design requires CSI, thus the channel estimation for the proposed transmissive RIS transceiver is essential. It is worth noting that the transmissive RIS passively transmits the signal and has no transmitting and receiving capabilities. For uplink transmission, because the receiving antenna has the receiving ability, the RIS transmissive element has no receiving ability, we can obtain the CSI of the uplink cascaded channel by through the direct cascaded channel estimation method proposed by \cite{9475488,9839429,8879620,8937491}, etc.. For ease of analysis, we temporarily assume that CSI is perfectly obtained. A more practical robust uplink algorithm based on imperfect CSI can be used as a further work.
\section{Joint Subcarrier Allocation, Power Allocation and RIS Transmissive Coefficient Design Algorithm}
In this section, in order to solve the formulated non-convex optimization problem (P0), we divide the problem (P0) into two sub-problems by applying the AO algorithm. Specifically, the first sub-problem is to fix the RIS transmissive coefficient and jointly optimize the power allocation and subcarrier allocation. The second sub-problem is to give power allocation and subcarrier allocation, and optimize the RIS transmissive coefficient. Finally, the two sub-problems are iterated alternately until convergence is achieved.
\newcounter{my2}
\begin{figure*}[!t]
	\normalsize
	\setcounter{my2}{\value{equation}}
	\setcounter{equation}{21}
	\begin{equation}
		\begin{aligned}
			{\cal L}\left( {{\bf{\tilde P}},{\bf{\tilde A}},{\bm{\lambda }},{\bm{\mu }}} \right) = \left( {\sum\limits_{k = 1}^K {\left( {W + {\mu _k}} \right)} } \right)\sum\limits_{n = 1}^N {{{\tilde a}_{k,n}}{{\log }_2}\left( {1 + \frac{{{{\tilde p}_{k,n}}{\Gamma _{k,n}}}}{{{{\tilde a}_{k,n}}}}} \right)}  - \sum\limits_{k = 1}^K {\sum\limits_{n = 1}^N {{\lambda _k}{{\tilde p}_{k,n}}} } + \sum\limits_{k = 1}^K {{\lambda _k}p_k^{\max }}  - \sum\limits_{k = 1}^K {{\mu _k}r_k^{\min }} .
		\end{aligned}
	\end{equation}
	\setcounter{equation}{\value{my2}}
	\hrulefill
	\vspace*{4pt}
\end{figure*}
\subsection{Problem Transformation}
The existence of the binary variable ${a_{k,n}}$ makes the problem (P0) a non-convex optimization problem. In order to solve the problem (P0), we first relax the binary variable ${a_{k,n}}$ to obtain ${\tilde a_{k,n}}$, i.e., ${a_{k,n}} \in \left\{ {0,1} \right\} \Rightarrow {\tilde a_{k,n}} \in \left[ {0,1} \right],\forall k,n$. Then, the auxiliary variable ${\tilde p_{k,n}} = {\tilde a_{k,n}}{p_{k,n}},\forall k,n$ is also introduced. After variable relaxation and the introduction of auxiliary variables, the optimization problem (P0) can be approximately transformed into the optimization problem (P1), which can be expressed as
\begin{subequations}
	\begin{align}
		\left( {{\textrm{P1}}} \right){\rm{~~~~~}}&\!\!\!\!\!\!\!\!\!\!\!\mathop {\max }\limits_{{\bf{\tilde A}},{\bf{\tilde P}},{\bf{c}}} {\rm{  }}W\sum\limits_{k = 1}^K {\sum\limits_{n = 1}^N {{{\tilde a}_{k,n}}{{\log }_2}\left(\! \!{1\! +\! \frac{{{{\tilde p}_{k,n}}{{\left| {{{\bf{h}}_n^H}{\rm{diag}}\left( {{{\bf{g}}_{k,n}}} \right){\bf{c}}} \right|}^2}{\nu _k}}}{{{{\tilde a}_{k,n}}{N_0}W}}} \!\!\right)} }, \nonumber  \\
		\rm{s.t.}\qquad &{{\tilde p}_{k,n}} \ge 0,\forall k,n,\\
		&\sum\limits_{n = 1}^N {{{\tilde p}_{k,n}}}  \le p_k^{\max },\forall k,\\
		&{{\tilde a}_{k,n}} \in \left[ {0,1} \right],\forall k,n,\\
		&\sum\limits_{k = 1}^K {{{\tilde a}_{k,n}} \le 1,} \forall n,\\
		&\!\!\!\!\!\!\!\!\!\!\!\sum\limits_{n = 1}^N {{{\tilde a}_{k,n}}{{\log }_2}\left(\!\! {1 \!+\! \frac{{{{\tilde p}_{k,n}}{{\left| {{{\bf{h}}_n^H}{\rm{diag}}\left( {{{\bf{g}}_{k,n}}} \right){\bf{c}}} \right|}^2}{\nu _k}}}{{{{\tilde a}_{k,n}}{N_0}W}}} \!\right)} \! \ge\! r_k^{\min },\!\forall k,\\
		&\left| {{c_m}} \right| \le 1,\forall m.
	\end{align}
\end{subequations}
Although we relax the binary variables, the three optimization variables are still highly coupled, which makes the optimization problem (P1) still a non-convex optimization problem. In the following, we use the AO algorithm framework to decouple the three coupled optimization variables, and then divide the problem (P1) into two sub-problems to solve $\bf{P}$, $\bf{A}$ and $\bf{c}$, respectively.

\subsection{Joint Optimization of Multi-user Power Allocation and Subcarrier Allocation}
In this sub-section, we first fix the RIS transmissive coefficient $\bf{c}$ to solve the user's power allocation $\bf{P}$ and subcarrier allocation $\bf{A}$. Let ${\Gamma _{k,n}} = \frac{{{{\left| {{{\bf{h}}_n^H}{\rm{diag}}\left( {{{\bf{g}}_{k,n}}} \right){\bf{c}}} \right|}^2}{\nu _k}}}{{{N_0}W}},\forall k,n$, the problem (P1) can be transformed into the problem (P2), which can be expressed as
\begin{subequations}
	\begin{align}
		\left( {{\textrm{P2}}} \right){\rm{~~~~~}}&\mathop {\max }\limits_{{\bf{\tilde A}},{\bf{\tilde P}}} {\rm{~~  }}W\sum\limits_{k = 1}^K {\sum\limits_{n = 1}^N {{{\tilde a}_{k,n}}{{\log }_2}\left( {1 + \frac{{{{\tilde p}_{k,n}}{\Gamma _{k,n}}}}{{{{\tilde a}_{k,n}}}}} \right)} },  \nonumber \\
		\rm{s.t.}\qquad &{{\tilde p}_{k,n}} \ge 0,\forall k,n,\\
		&\sum\limits_{n = 1}^N {{{\tilde p}_{k,n}}}  \le p_k^{\max },\forall k,\\
		&{{\tilde a}_{k,n}} \in \left[ {0,1} \right],\forall k,n,\\
		&\sum\limits_{k = 1}^K {{{\tilde a}_{k,n}} \le 1,} \forall n,\\
		&\sum\limits_{n = 1}^N {{{\tilde a}_{k,n}}{{\log }_2}\left( {1 + \frac{{{{\tilde p}_{k,n}}{\Gamma _{k,n}}}}{{{{\tilde a}_{k,n}}}}} \right)}  \ge r_k^{\min },\forall k.
	\end{align}
\end{subequations}

\emph{Theorem 1: The function $h\left( {x,t} \right) = t{\log _2}\left( {1 + \frac{x}{t}} \right)$ is concave with respect to (w.r.t) $x>0$ and $t>0$.}

\emph{Proof:} The function $h\left( {x,t} \right)$ can be obtained by the perspective transformation of the function $f\left( x \right) = {\log _2}\left( {1 + x} \right)$, i.e., $h\left( {x,t} \right) = tf\left( {\frac{x}{t}} \right)$. Since the perspective function is concave-preserving and $f\left( x \right) = {\log _2}\left( {1 + x} \right)$ is a concave function w.r.t $x>0$, the function $h\left( {x,t} \right) = t{\log _2}\left( {1 + \frac{x}{t}} \right)$ is a concave function w.r.t $x>0$ and $t>0$. The proof of \emph{Theorem 1} is completed. $\hfill\blacksquare$

According to \emph{Theorem 1}, the objective function in the optimization problem (P2) is the sum of several concave functions, so it is concave w.r.t ${{{\tilde a}_{k,n}}}$ and ${{{\tilde p}_{k,n}}}$. In addition, constraints (17a)-(17d) are affine, and constraint (17e) is convex, so this problem is a standard convex optimization problem. Herein, we apply the Lagrangian dual decomposition technique to solve the convex optimization problem (P2). Define ${\bm{\lambda }} = \left[ {{\lambda _1},...,{\lambda _K}} \right]$ and ${\bm{\mu }} = \left[ {{\mu _1},...,{\mu _K}} \right]$ as the Lagrangian multipliers corresponding to constraint (17b) and constraint (17e), respectively. The Lagrangian function of problem (P2) can be written as Eq. (18). It is worth noting that other constraints will be considered in the later solving section, so they will not be ignored. The Lagrangian dual function can be expressed as
\setcounter{equation}{18}
\begin{equation}
	g\left( {{\bm{\lambda }},{\bm{\mu }}} \right) = \mathop {\sup }\limits_{{\bf{\tilde P}},{\bf{\tilde A}}} {\cal L}\left( {{\bf{\tilde P}},{\bf{\tilde A}},{\bm{\lambda }},{\bm{\mu }}} \right).
\end{equation}
Therefore, the Lagrangian dual problem can be given by
\begin{subequations}
	\begin{align}
		\min {\rm{~~~~~~~}}g\left( {{\bm{\lambda }},{\bm{\mu }}} \right), \nonumber \\
		\rm{s.t.}\qquad {\bm{\lambda }},{\bm{\mu }} \succeq 0.
	\end{align}
\end{subequations}

In each iteration of the Lagrangian dual decomposition method, Lagrangian multipliers are first given to solve a maximization problem by jointly optimizing power allocation and subcarrier allocation. Then, the obtained power allocation and subcarrier allocation schemes are applied to solve a minimization problem of Lagrangian multipliers. Finally, the two problem are iterated alternately until Lagrangian multipliers convergence is achieved. In the following, the Lagrangian dual decomposition method will be elaborated in detail.

First, in the $i$-th iteration, given the Lagrangian multipliers $\bm{\lambda }$ and $\bm{\mu }$, we solve an unconstrained maximization problem, which can be expressed as
\begin{equation}
	{{\bf{\tilde P}}^i},{{\bf{\tilde A}}^i} = \mathop {\arg \max }\limits_{{\bf{\tilde P}},{\bf{\tilde A}}} {\cal L}\left( {{\bf{\tilde P}},{\bf{\tilde A}},{\bm{\lambda }},{\bm{\mu }}} \right),
\end{equation}
where ${{\bf{\tilde P}}^i}$ and ${{\bf{\tilde A}}^i}$ are the subcarrier allocation and power allocation schemes obtained in the $i$-th iteration, respectively.

\emph{Theorem 2: ${\cal L}\left( {{\bf{\tilde P}},{\bf{\tilde A}},{\bm{\lambda }},{\bm{\mu }}} \right)$ of Eq. (18) is a concave function w.r.t ${{{\tilde a}_{k,n}}}$ and ${{{\tilde p}_{k,n}}}$.}

\emph{Proof:} ${\cal L}\left( {{\bf{\tilde P}},{\bf{\tilde A}},{\bm{\lambda }},{\bm{\mu }}} \right)$ of Eq. (18) can be written as Eq. (22). Given the Lagrange multipliers $\bm{\lambda }$ and $\bm{\mu }$, according to \emph{Theorem 1}, the first term on the right-hand-side (RHS) of the function is concave w.r.t ${{{\tilde a}_{k,n}}}$ and ${{{\tilde p}_{k,n}}}$, and the last three term on the RHS are affine w.r.t ${{{\tilde a}_{k,n}}}$ and ${{{\tilde p}_{k,n}}}$. Therefore, ${\cal L}\left( {{\bf{\tilde P}},{\bf{\tilde A}},{\bm{\lambda }},{\bm{\mu }}} \right)$ of Eq. (18) is a concave function w.r.t ${{{\tilde a}_{k,n}}}$ and ${{{\tilde p}_{k,n}}}$. The proof of \emph{Theorem 2} is completed. $\hfill\blacksquare$

Since the problem (P2) is a convex problem, the solution that satisfies the Karush-Kuhn-Tucker (KKT) condition is also the optimal solution of the problem (P2) and its dual problem \cite{boyd2004convex}. Therefore, we use the gradient condition in the KKT condition to solve the user's power allocation and subcarrier allocation. Since ${\cal L}\left( {{\bf{\tilde P}},{\bf{\tilde A}},{\bm{\lambda }},{\bm{\mu }}} \right)$ is concave to w.r.t ${{{\tilde a}_{k,n}}}$ and ${{{\tilde p}_{k,n}}}$, we take the partial derivation of ${\cal L}\left( {{\bf{\tilde P}},{\bf{\tilde A}},{\bm{\lambda }},{\bm{\mu }}} \right)$ to ${{{\tilde p}_{k,n}}}$, and the multi-user power allocation scheme can be obtained as follows:
\setcounter{equation}{22}
\begin{equation}
	p_{k,n}^{ * ,i} = {\left[ {\frac{{W + {\mu _k}}}{{{\lambda _k}\ln 2}} - \frac{1}{{{\Gamma _{k,n}}}}} \right]^ + },\forall k,n,
\end{equation}
where ${\left[  \cdot  \right]^ + }$ returens the maximum value of $(\cdot)$ and zero. It can be seen that when the RIS transmissive coefficient is given, the uplink users power allocation mainly depends on different Lagrangian multipliers.

In addition, in order to obtain the subcarrier allocation scheme, we take the partial derivative of ${\cal L}\left( {{\bf{\tilde P}},{\bf{\tilde A}},{\bm{\lambda }},{\bm{\mu }}} \right)$ to ${{{\tilde a}_{k,n}}}$, and the subcarrier allocation criterion can be obtained, denoted by $\chi _{k,n}^i$, which can be further expressed as
\begin{equation}
	\begin{aligned}
		\chi _{k,n}^i = \left( {W + {\mu _k}} \right)\left( {{{\log }_2}\left( {1 + p_{k,n}^{ * ,i}{\Gamma _{k,n}}} \right)} -\right.\\
			\left. {\frac{{p_{k,n}^{ * ,i}{\Gamma _{k,n}}}}{{\left( {1 + p_{k,n}^{ * ,i}{\Gamma _{k,n}}} \right)\ln 2}}} \right),\forall k,n.
	\end{aligned}
\end{equation}
This subcarrier allocation criterion can be regarded as the sum-rate growth rate when the $n$-th subcarrier is allocated to the $k$-th user. Meanwhile, for $p_{k,n}^{ * ,i} > 0$, $\chi _{k,n}^i > 0$ means that when the $n$-th subcarrier is allocated to the $k$-th user, the system sum-rate is guaranteed not to be reduced. We can allocate subcarriers for users according to the subcarrier allocation criterion. Specifically, in order to maximize the sum-rate of the system, the $n$-th subcarrier is allocated to the user with the largest $\chi _{k,n}^i$, which can be expressed as
\begin{equation}
	a_{k,n}^{ * ,i} = \left\{ {\begin{array}{*{20}{c}}
			{1,{\rm{ ~~if  ~}}\chi _{k,n}^i = \mathop {\mathop {\max }\limits_k \left( {\chi _{k,n}^i} \right){\rm{ ~and  ~ }}\chi _{k,n}^i > 0,}\limits_{} }\\
			{0,{\rm{   ~~ otherwise}}{\rm{.~~~~~~~ ~~~~~~~~~~~~~~~~~~~~~~}}}
	\end{array}} \right.
\end{equation}

Since ${\cal L}\left( {{\bf{\tilde P}},{\bf{\tilde A}},{\bm{\lambda }},{\bm{\mu }}} \right)$ is differentiable and the solution of problem (21) is unique, the Lagrangian multipliers $\bm{\lambda }$ and $\bm{\mu }$ can be obtained by the method of gradient update \cite{6514978}. In the $i$-th iteration, the Lagrangian multipliers can be updated by
\begin{equation}
	\lambda _k^i = \left[ {\lambda _k^{i - 1} - {\varsigma _k}\left( {p_k^{\max } - \sum\limits_{n = 1}^N {a_{k,n}^{ * ,i}p_{k,n}^{ * ,i}} } \right)} \right]^+,
\end{equation}
and
\begin{equation}
	\mu _k^i = \left[ {\mu _k^{i - 1} - {\psi _k}\left( {\sum\limits_{n = 1}^N {a_{k,n}^{ * ,i}{{\log }_2}\left( {1 + p_{k,n}^{ * ,i}{\Gamma _{k,n}}} \right) - r_k^{\min }} } \right)} \right]^+,
\end{equation}
where ${\varsigma _k}$ and ${\psi _k}$ are the update stepsizes of the corresponding multipliers, which are used to control the convergence of the Lagrangian multipliers. Accordingly, we can obtain power allocation and subcarrier allocation schemes with a fixed RIS transmissive coefficient. The power allocation and subcarrier allocation algorithm can be shown as $\textbf{Algorithm 1}$.
\begin{algorithm}[H]
	\caption{The Power Allocation and Subcarrier Allocation Algorithm} 
	\begin{algorithmic}[1]
		\State $\textbf{Input:}$ ${{\bf{A}}^0}$, ${{\bf{P}}^0}$, ${{\bf{c}}^0}$, ${{\bm{\lambda }}^0} $, ${{\bm{\mu }}^0}$, convergence threshold $\epsilon$ and iteration index $i = 0$.
		\Repeat 
		\State The power allocation scheme can be obtained with Eq. (23).
		\State The subcarrier allocation scheme can be obtained with Eq. (24) and Eq. (25).
		\State Update Lagrangian multipliers $\bm{\lambda }$ and $\bm{\mu }$ with Eq. (26) and Eq. (27).
		\State Update $i = i + 1$.
		\Until the Lagrangian multipliers meets the convergence threshold $\epsilon$.
		\State $\textbf{Output:}$ power allocation ${{\bf{P}}^ * }$ and subcarrier allocation ${{\bf{A}}^ * }$.
	\end{algorithmic}
\end{algorithm}

\subsection{Optimization of RIS Transmissive Coefficient}
In this subsection, given the subcarrier allocation matrix $\bf{A}$ and the user's power allocation $\bf{P}$, we solve the RIS transmissive coefficient. Let ${\Xi _{k,n}} = \frac{{{p_{k,n}}{\nu _k}}}{{{N_0}W}},\forall k,n$, the problem (P1) can be transformed into the problem (P3), which can be given by
\begin{subequations}
	\begin{align}
		\left( {{\textrm{P3}}} \right){\rm{~~~~~}}&\!\!\!\!\!\!\!\!\!\mathop {\max }\limits_{\bf{c}} {\rm{~}}W\sum\limits_{k = 1}^K {\sum\limits_{n = 1}^N {{{\tilde a}_{k,n}}{{\log }_2}\left( {1 + {\Xi _{k,n}}{{\left| {{{\bf{h}}_n^H}{\rm{diag}}\left( {{{\bf{g}}_{k,n}}} \right){\bf{c}}} \right|}^2}} \right)} } ,  \nonumber\\
		\rm{s.t.}\qquad &\!\!\!\!\!\!\!\!\!\sum\limits_{n = 1}^N {{{\tilde a}_{k,n}}W{{\log }_2}\left(\! {1 \!+\! {\Xi _{k,n}}{{\left| {{{\bf{h}}_n^H}{\rm{diag}}\left( {{{\bf{g}}_{k,n}}} \right){\bf{c}}} \right|}^2}} \right)} \! \ge\! r_k^{\min },\forall k,\\
		&\left| {{c_m}} \right| \le 1,\forall m.
	\end{align}
\end{subequations}
Let ${\bf{v}}_{k,n}^H = {{\bf{h}}_n^H}{\rm{diag}}\left( {{{\bf{g}}_{k,n}}} \right) \in {\mathbb{C}^{1 \times M}}$, ${\left| {{{\bf{h}}_n^H}{\rm{diag}}\left( {{{\bf{g}}_{k,n}}} \right){\bf{c}}} \right|^2} = {\left| {{\bf{v}}_{k,n}^H{\bf{c}}} \right|^2} = {\bf{v}}_{k,n}^H{\bf{c}}{{\bf{c}}^H}{{\bf{v}}_{k,n}}$. Herein, let ${\bf{C}} = {\bf{c}}{{\bf{c}}^H} \in {\mathbb{C}^{M \times M}}$, where ${\bf{C}} \succeq 0$ and ${\rm{rank}}\left( {\bf{C}} \right) = 1$. In addition, let ${{\bf{V}}_{k,n}} = {{\bf{v}}_{k,n}}{\bf{v}}_{k,n}^H \in {\mathbb{C}^{M \times M}}$, then ${\left| {{{\bf{h}}_n^H}{\rm{diag}}\left( {{{\bf{g}}_{k,n}}} \right){\bf{c}}} \right|^2} = {\rm{tr}}\left( {{\bf{C}}{{\bf{V}}_{k,n}}} \right)$. The problem (P3) is equivalently expressed as the problem (P4), which can be expressed as
\begin{subequations}
	\begin{align}
		\left( {{\textrm{P4}}} \right){\rm{~~~~~}}&\mathop {\max }\limits_{\bf{C}} {\rm{~~~}}W\sum\limits_{k = 1}^K {\sum\limits_{n = 1}^N {{{\tilde a}_{k,n}}{{\log }_2}\left( {1 + {\Xi _{k,n}}{\rm{tr}}\left( {{\bf{C}}{{\bf{V}}_{k,n}}} \right)} \right)} },  \nonumber  \\
		\rm{s.t.}\qquad &{\rm{ }}\sum\limits_{n = 1}^N {{{\tilde a}_{k,n}}W{{\log }_2}\left( {1 + {\Xi _{k,n}}{\rm{tr}}\left( {{\bf{C}}{{\bf{V}}_{k,n}}} \right)} \right)}  \ge r_k^{\min },\forall k,\\
		&{{\bf{C}}_{m,m}} \le 1,\forall m,\\
		&{\bf{C}} \succeq 0,\\
		&{\rm{rank}}\left( {\bf{C}} \right) = 1.
	\end{align}
\end{subequations}
Due to the existence of the non-convex rank-one constraint (29d), the problem (P4) is still a non-convex optimization problem. We apply \emph{Proposition 1} to transform the constraint (29d).

\emph{Proposition 1: For the positive semi-definite matrix ${\bf{B}} \in {\mathbb{C}^{N \times N}}$, ${\rm{tr}}\left( {\bf{B}} \right) > 0$, the rank-one constraint can be equivalent to the difference between two convex functions, which can be given by
\begin{equation}
	{\rm{rank}}\left( {\bm{{\rm B}}} \right) = 1 \Rightarrow {\rm{tr}}\left( {\bm{{\rm B}}} \right) - {\left\| {\bm{{\rm B}}} \right\|_2},
\end{equation}
where ${\rm{tr}}\left( {\bf{B}} \right) = \sum\limits_{n = 1}^N {{\sigma _n}\left( {\bf{B}} \right)} $, ${\left\| {\bf{B}} \right\|_2} = {\sigma _1}\left( {\bf{B}} \right)$ is spectral norm, and ${\sigma _n}\left( {\bf{B}} \right)$ represents the $n$-th largest singular value of matrix ${\bf{B}}$.} 

According to \emph{Proposition 1}, we can transform the constraint (29d) in the problem (P4) into
\begin{equation}
	{\rm{rank}}\left( {\bf{C}} \right){\rm{ = 1}} \Rightarrow {\rm{tr}}\left( {\bf{C}} \right) - {\left\| {\bf{C}} \right\|_2}.
\end{equation}
We introduce the penalty factor $\xi $, and add Eq. (31) to the objective function of the problem (P4). The problem (P4) can be transformed into the problem (P5), which can be given by
\begin{subequations}
	\begin{align}
		\left( {{\textrm{P5}}} \right){\rm{~~~~~}}&\mathop {\max }\limits_{\bf{C}} {\rm{ ~~~}}W\sum\limits_{k = 1}^K {\sum\limits_{n = 1}^N {{{\tilde a}_{k,n}}{{\log }_2}\left( {1 + {\Xi _{k,n}}{\rm{tr}}\left( {{\bf{C}}{{\bf{V}}_{k,n}}} \right)} \right)} }  \nonumber \\ &~~~~~~~~~~- \xi \left( {{\rm{tr}}\left( {\bf{C}} \right) - {{\left\| {\bf{C}} \right\|}_2}} \right),  \nonumber \\
		\rm{s.t.}\qquad &{\rm{ }}\sum\limits_{n = 1}^N {{{\tilde a}_{k,n}}W{{\log }_2}\left( {1 + {\Xi _{k,n}}{\rm{tr}}\left( {{\bf{C}}{{\bf{V}}_{k,n}}} \right)} \right)}  \ge r_k^{\min },\forall k,\\
		&{{\bf{C}}_{m,m}} \le 1,\forall m,\\
		&{\bf{C}} \succeq 0.
	\end{align}
\end{subequations}
Since ${\left\| {\bf{C}} \right\|_2}$ is convex, the problem (P5) is a difference-convex (DC) programming problem, which is still a non-convex optimization problem. In this paper, we apply SCA to obtain the lower bound of ${\left\| {\bf{C}} \right\|_2}$, which can be expressed as
\begin{equation}
	\begin{aligned}
		{\left\| {\bf{C}} \right\|_2} \!\ge\! {\left\| {{{\bf{C}}^{\left( t \right)}}} \right\|_2} \!+\! {\rm{tr}}\left(\! {{{\bf{u}}_{\max }}\left(\! {{{\bf{C}}^{\left( t \right)}}} \right){{\bf{u}}_{\max }}{{\left(\! {{{\bf{C}}^{\left( t \right)}}} \right)}^H}\!\left( {{\bf{C}}\! -\! {{\bf{C}}^{\left( t \right)}}}\! \right)} \!\right)\\ \buildrel \Delta \over = {\left( {{{\left\| {\bf{C}} \right\|}_2}} \right)^{lb}},
	\end{aligned}
\end{equation}
where ${{\bf{u}}_{\max }}\left( {{{\bf{C}}^{\left( t \right)}}} \right)$ represents the eigenvector corresponding to the largest singular value of matrix ${{\bf{C}}^{\left( t \right)}}$. Thus, the problem (P5) can be transformed into the problem (P6) as follows
\begin{subequations}
	\begin{align}
		\left( {{\textrm{P6}}} \right){\rm{~~~~~}}&\mathop {\max }\limits_{\bf{C}} {\rm{ ~~~}}W\sum\limits_{k = 1}^K {\sum\limits_{n = 1}^N {{{\tilde a}_{k,n}}{{\log }_2}\left( {1 + {\Xi _{k,n}}{\rm{tr}}\left( {{\bf{C}}{{\bf{V}}_{k,n}}} \right)} \right)} }  \nonumber \\ &~~~~~~~~~~- \xi \left( {{\rm{tr}}\left( {\bf{C}} \right) - {{\left( {{{\left\| {\bf{C}} \right\|}_2}} \right)}^{lb}}} \right) ,\nonumber \\
		\rm{s.t.}\qquad &{\rm{ }}\sum\limits_{n = 1}^N {{{\tilde a}_{k,n}}W{{\log }_2}\left( {1 + {\Xi _{k,n}}{\rm{tr}}\left( {{\bf{C}}{{\bf{V}}_{k,n}}} \right)} \right)}  \ge r_k^{\min },\forall k,\\
		&{{\bf{C}}_{m,m}} \le 1,\forall m,\\
		&{\bf{C}} \succeq 0.
	\end{align}
\end{subequations}
It is easy to find that the problem (P6) is a standard SDP problem, which can be solved by using the CVX toolbox to obtain the design of the RIS transmissive coefficient \cite{grant2014cvx}. 

\subsection{The Overall Power Allocation, Subcarrier Allocation and RIS Transmissive Coefficient Design Algorithm}
In this subsection, we propose the overall power allocation, subcarrier allocation and RIS transmissive coefficient design algorithm, which can be summarized as $\textbf{Algorithm 2}$. First, the power allocation and subcarrier allocation are determined by applying $\textbf{Algorithm 1}$. Then, the RIS transmissive coefficient design can be obtained by solving the problem (P6). Finally, two sub-problems are alternately solved to achieve convergence.
\subsection{Computational Complexity and Convergence Analysis}
\subsubsection{Computational complexity analysis}
The complexity of $\textbf{Algorithm 1}$ is ${\cal O}\left( {KN} \right)$. In addition, the complexity of solving SDP problems (P6) by the interior point method can be denoted by ${\cal O}\left( {(M)^{3.5}} \right)$ in each iteration \cite{boyd2004convex}. While the subgradient can be computed by singular value decomposition (SVD) with complexity ${\cal O}\left( {{(M)^{3}}} \right)$. Therefore, the computational complexity is at most ${\cal O}\left( {(M)^{3.5}} \right)$. In summary, let $t$ be the number of iterations required for the proposed algorithm to achieve convergence, the computational complexity of  $\textbf{Algorithm 2}$ can be denoted by ${\cal O}\left( {t\left( {KN + {(M)^{3.5}}} \right)} \right)$.

\subsubsection{Convergence analysis}
The convergence of the proposed $\textbf{Algorithm 2}$ for uplink transmissive RIS multi-antenna system with OFDMA can be proved as follows. 

We define ${{\bf{A}}^t}$, ${{\bf{P}}^t}$ and ${{\bf{c}}^t}$ as the $t$-th iteration solution of the problem (P2) and (P6). Herein, the objective function is denoted by ${\cal R}\left( {{{\bf{A}}^t},{{\bf{P}}^t},{{\bf{c}}^t}} \right)$. In the step 3 of $\textbf{Algorithm 2}$, since the power allocation and subcarrier allocation can be obtained for given ${{\bf{c}}^t}$. Hence, we have 
\begin{equation}
	{\cal R}\left( {{{\bf{A}}^t},{{\bf{P}}^t},{{\bf{c}}^t}} \right) \le {\cal R}\left( {{{\bf{A}}^{t + 1}},{{\bf{P}}^{t + 1}},{{\bf{c}}^t}} \right).
\end{equation}
In the step 4 of $\textbf{Algorithm 2}$, the RIS transmissive coefficient can be obtained when ${{\bf{A}}^t}$ and ${{\bf{P}}^t}$  are given. Herein, we also have 
\begin{equation}
	{\cal R}\left( {{{\bf{A}}^{t + 1}},{{\bf{P}}^{t + 1}},{{\bf{c}}^t}} \right) \le {\cal R}\left( {{{\bf{A}}^{t + 1}},{{\bf{P}}^{t + 1}},{{\bf{c}}^{t + 1}}} \right).
\end{equation}
Based on the above, we can obtain
\begin{equation}
	{\cal R}\left( {{{\bf{A}}^t},{{\bf{P}}^t},{{\bf{c}}^t}} \right) \le {\cal R}\left( {{{\bf{A}}^{t + 1}},{{\bf{P}}^{t + 1}},{{\bf{c}}^{t + 1}}} \right).
\end{equation}
It shows that the value of the objective function after each iteration of $\textbf{Algorithm 2}$ is non-decreasing. Meanwhile, the objective function value of the problem (P1) has an upper bound, so the convergence of $\textbf{Algorithm 2}$ can be guaranteed.
\begin{algorithm}[H]
	\caption{The Overall Power Allocation, Subcarrier Allocation and RIS Transmissive Coefficient Design Algorithm} 
	\begin{algorithmic}[1]
		\State$\textbf{Input:}$ ${{\bf{A}}^0}$, ${{\bf{P}}n^0}$, ${{\bf{c}}^0}$, ${{\bm{\lambda }}^0} $, ${{\bm{\mu }}^0}$, convergence threshold $\epsilon$ and iteration index $t = 0$.
		\Repeat
		\State Obtain power allocation ${\bf{P}}^ * $ and subcarrier allocation ${\bf{A}}^ * $ according to the $\textbf{Algorithm 1}$.
		\State Obtain RIS transmissive coefficient ${\bf{c}}^ * $ by solving the problem (P6).
		\State Update $t=t+1$.
		\Until The fractional decrease of the objective value is below a threshold $\epsilon$.
		\State \Return The power allocation, subcarrier allocation and RIS transmissive coefficient.
	\end{algorithmic}
\end{algorithm}
\section{Numerical results}
In this section, we elaborate the effectiveness of the proposed joint resource allocation and RIS transmissive coefficient design algorithm in transmissive RIS transceiver enabled uplink communication system with OFDMA via numerical simulations. The considered simulation setup of the system is depicted in Fig. 5. We consider a three-dimensional coordinate system in this paper, where the location of RIS transceiver is (0m, 0m, 15m), the distance between receiving antenna and RIS is 20cm, and $K = 5$ users are randomly distributed in a square horizontal plane with a side length of 50m, regardless of the user's height. Moreover, the receiving antenna is equipped with single antenna, and transmissive RIS is equipped with $M=25$ transmissive elements. We set the antenna spacing to be half of the carrier wavelength, i.e., $d_r = \lambda/2$ and $d_c = \lambda/2$. Meanwhile, we set $P_k^{\max}=200$mW, ${\rm{BE}}{{\rm{R}}_k}=-30$dB, $N_0=-174$dBm/Hz, $f_c=3$GHz and $r_k^{min}=10$bps in our numerical simulations. In addition, the path loss exponent is set as $\alpha=3$. The path loss with a reference distance of 1m is set to $C_0=-30$dB, and $\rho$ is randomly generated from a distribution that obeys $ {\cal C}{\cal N}\left(0,1 \right)$.  We set Rician factor $\kappa$ to 3dB, and we set the convergence threshold of the proposed algorithm to $10^{-3}$. 
\begin{figure}
	\centerline{\includegraphics[width=8cm]{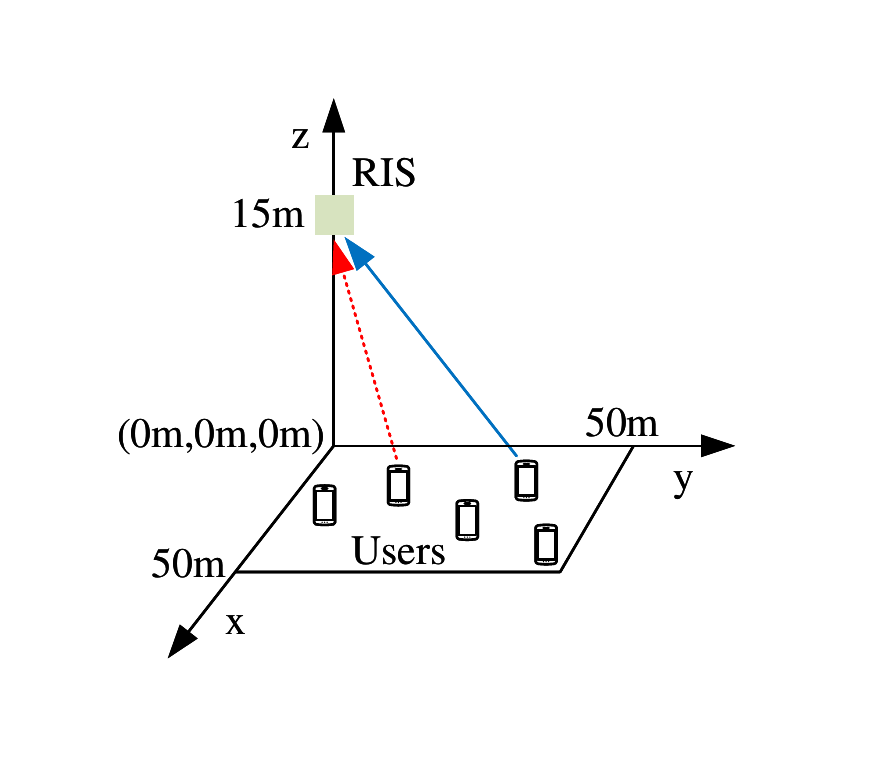}}
	\caption{Simulation setup of transmissive RIS transceiver enabled uplink communication system.}
\end{figure}

First, we verify the convergence of the proposed subcarrier allocation, power allocation and RIS transmissive coefficient design algorithm. Fig. 6 depicts the changes of the system sum-rate with the number of iterations under different maximum transmit powers of user. We can see that the system sum-rate increase with the increase in the number of iterations. At the same time, the proposed algorithm has a very good convergence performance, i.e., it can quickly achieve convergence. In addition, we can also compare the sum-rate relationship of users with different maximum transmit powers. It can be seen that the greater the maximum transmit power of the user, the greater the sum-rate of the system. This is because when the user's maximum transmit power increases, the user's allocated power will increase, and each user's uplink rate will increase, so the system sum-rate will increase.
\begin{figure}
	\centerline{\includegraphics[width=8cm]{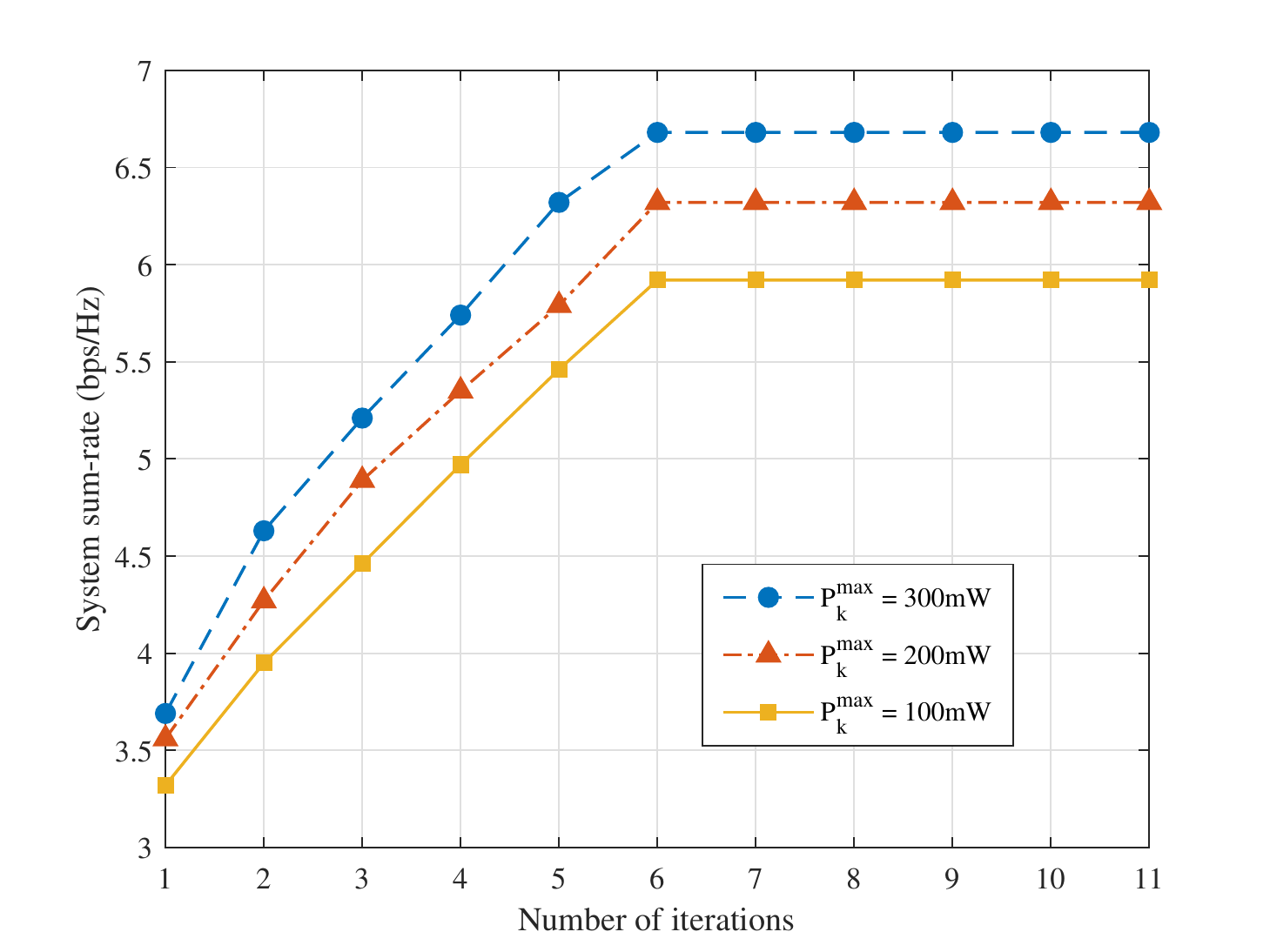}}
	\caption{The convergence of the proposed joint subcarrier allocation, power allocation and RIS transmissive coefficient design algorithm.}
\end{figure}

In order to evaluate the performance of the proposed algorithm, we consider the following benchmark algorithm. Benchmark 1 (three-stage algorithm): The algorithm does not perform alternate iterations after optimizing subcarrier allocation, power allocation and RIS transmission coefficient once. Benchmark 2 (random coefficient algorithm): This algorithm still uses the problem (P2) solution for subcarrier allocation and power allocation, and uses a random scheme for the RIS transmissive coefficient. Benchmark 3 (random allocation algorithm): This algorithm uses a random scheme for subcarrier allocation, power allocation, and RIS transmissive coefficient design.

In Fig. 7, when the user's maximum transmit power changes, we compare the performance of the proposed algorithm with different benchmark algorithms in term of system sum-rate. As observed, as the user's maximum transmit power increases, the system sum-rate under different algorithms increases. This is mainly due to the increase in the maximum transmit power of the user, which makes the power allocated by each user larger, and therefore the uplink sum-rate of the system becomes larger. In addition, the performance of our proposed algorithm is better than that of other benchmark algorithms. This is mainly due to the fact that benchmark 1 did not achieve global convergence. The RIS coefficient design of benchmark 2 cannot enhance the signal transmitted by the user in the uplink. The benchmark 3 scheme does not optimize the optimization variables, so the performance is poor. Finally, from the two sets of data corresponding to different user numbers $K=5$ and $K=10$, it can be seen that when the number of users increases, the system sum-rate will also increase.
\begin{figure}
	\centerline{\includegraphics[width=8cm]{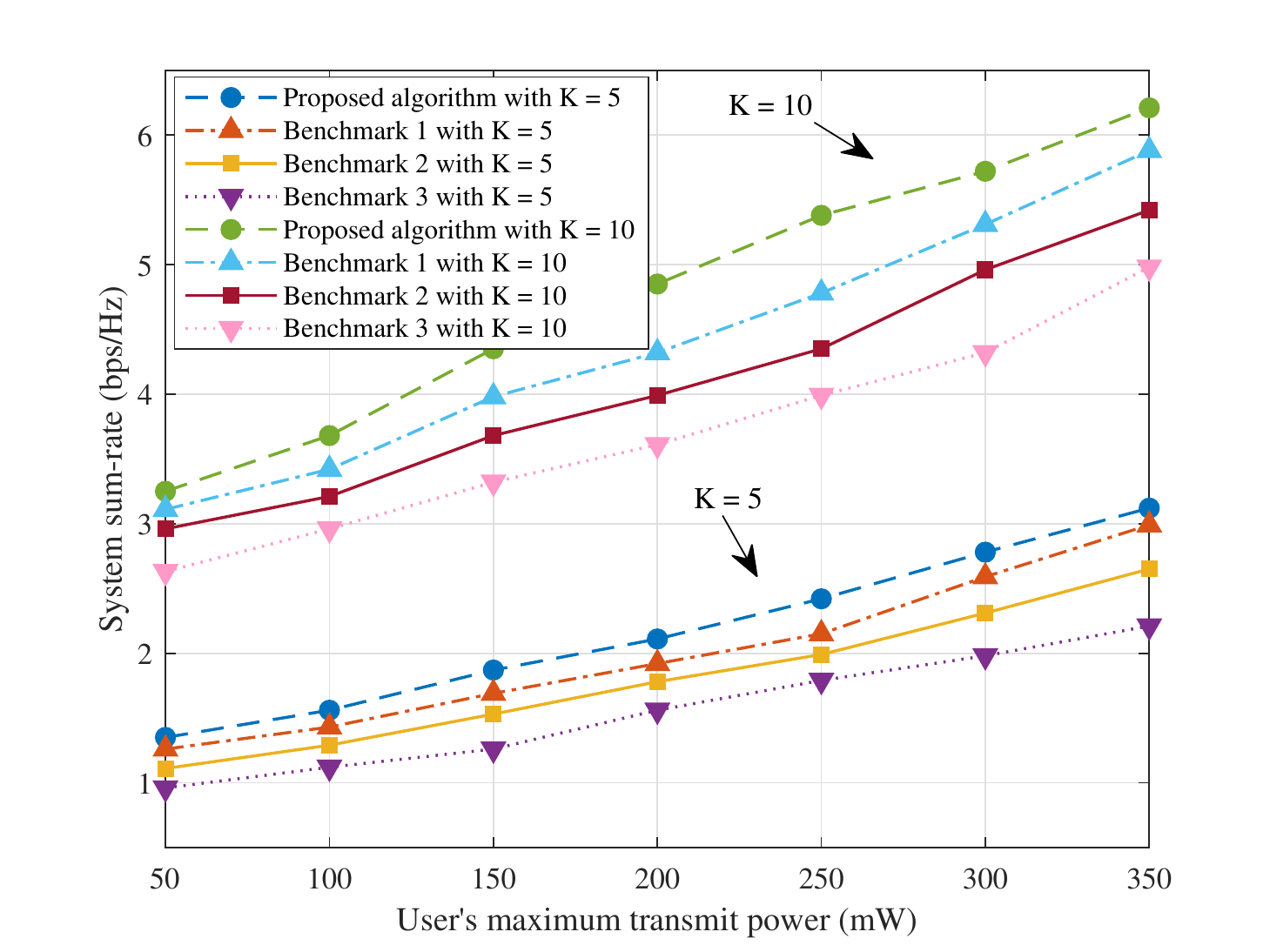}}
	\caption{System sum-rate versus the user's maximum transmit power with different $K$.}
\end{figure}

Next, Fig. 8 depicts the system sum-rate changes under different benchmark algorithms and different numbers of subcarriers when the user's maximum transmit power increases. As shown in the Fig. 8, when the user's maximum transmit power increases, the system sum-rate under different algorithms and different numbers of subcarriers all increase, and the reason is the same as mentioned above. Compared with other benchmark algorithms, the performance of the proposed algorithm is also superior. In addition, from the curves corresponding to different subcarriers $N=20$ and $N=40$, it can be seen that when the same algorithm is used, the more subcarriers, the greater the system sum-rate. This is because when the bandwidth is constant, the increase in the number of subcarriers can increase the frequency diversity gain, so the performance of the system in term of sum-rate can be improved.
\begin{figure}
	\centerline{\includegraphics[width=8cm]{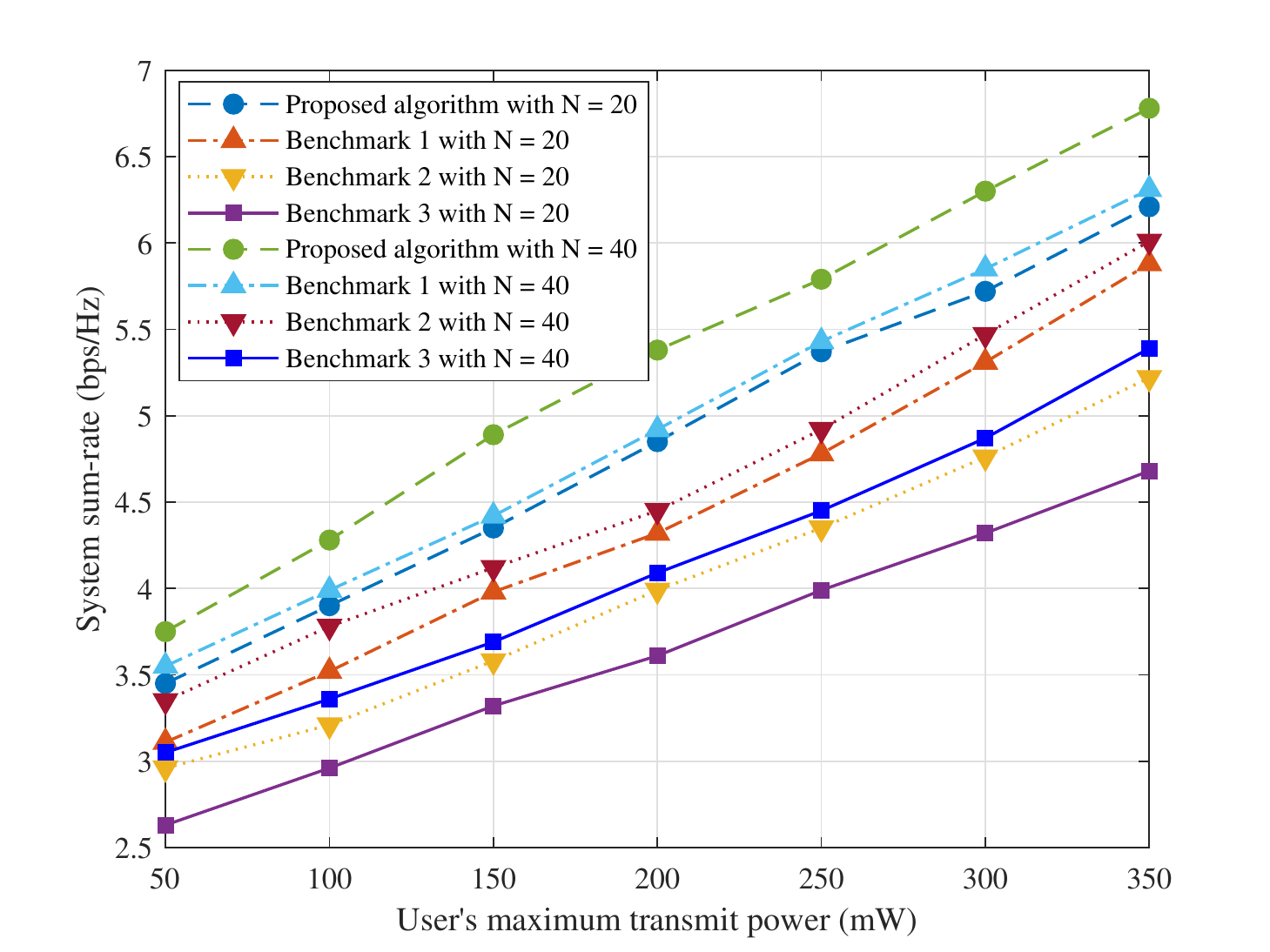}}
	\caption{System sum-rate versus the user's maximum transmit power with different $N$.}
\end{figure}

Fig. 9 shows the performance comparison between the proposed algorithm and different benchmark algorithms in term of system sum-rate when the number of users increases. As observed, when the number of users increases, the sum-rate of the system under different algorithms increases. This is mainly because the system sum-rate depends on the uplink transmission rate of all users, so when the number of users increases, the system sum-rate will increase. In addition, it can be seen that our proposed algorithm is still superior to other benchmark algorithms, the reason has been mentioned in the above discussion.
\begin{figure}
	\centerline{\includegraphics[width=8cm]{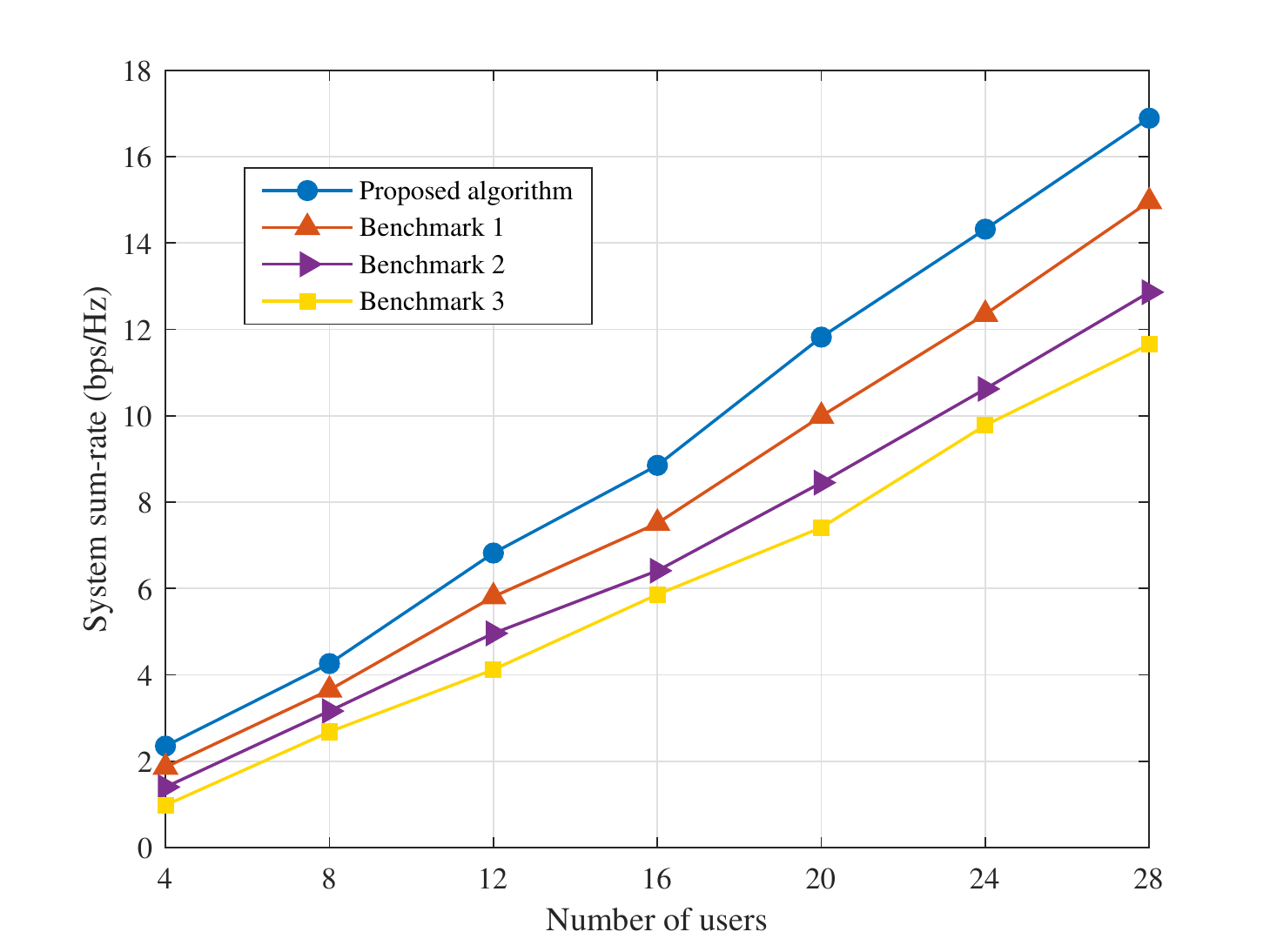}}
	\caption{System sum-rate versus number of users.}
\end{figure}

Then, we compare the variation of system sum-rate with the number of subcarriers under different algorithms, as shown in Fig. 10. It can be seen that as the number of subcarriers increases, the system sum-rate under different algorithms is constantly increasing. This is because we are considering an uplink OFDMA network based on transmissive RIS, where the bandwidth is certain, so the more subcarriers, the more frequency diversity gain it brings, so the system performance in term of sum-rate is superior. Then, the performance of our proposed algorithm is still better than other benchmark algorithms.
\begin{figure}
	\centerline{\includegraphics[width=8cm]{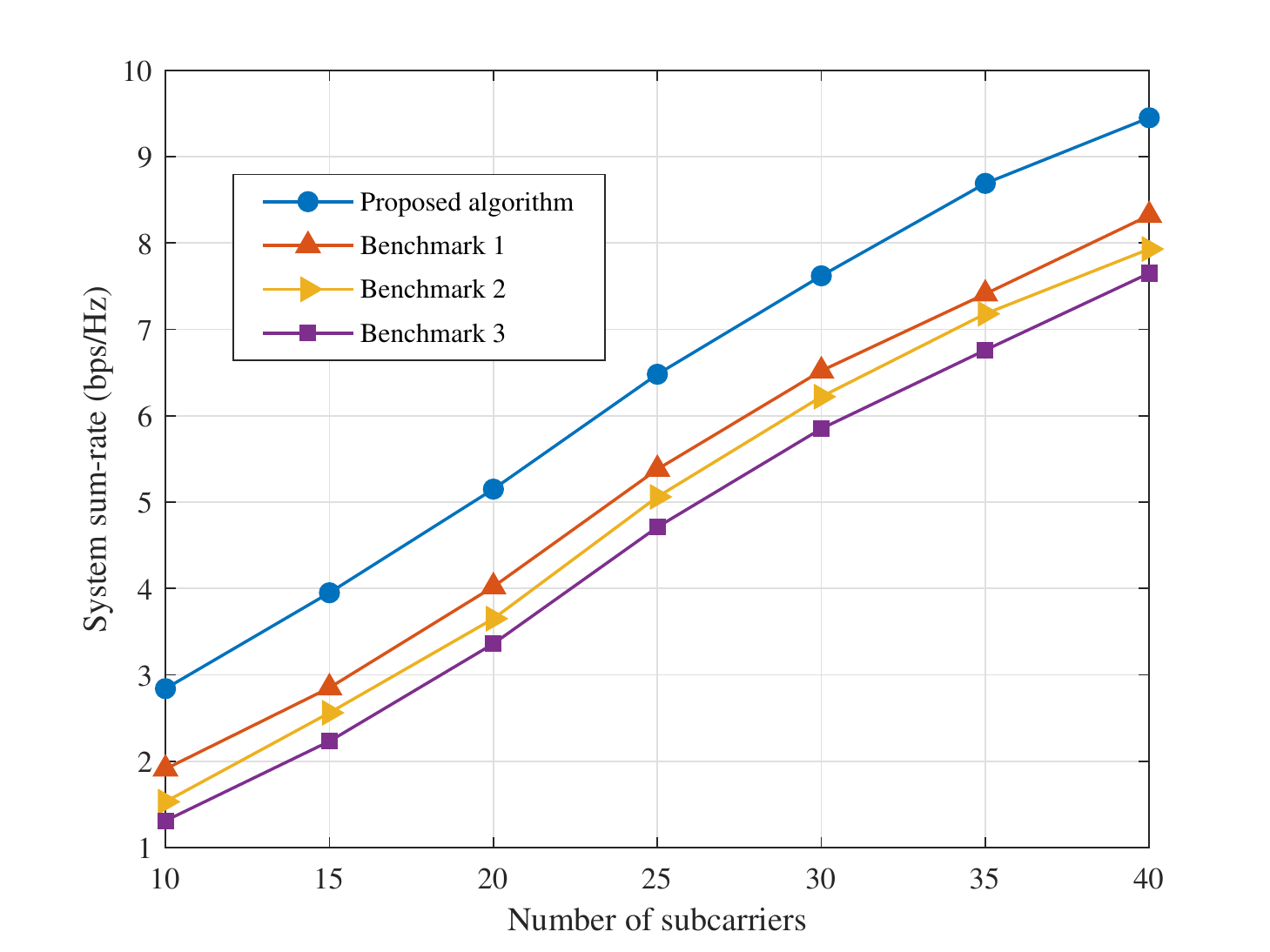}}
	\caption{System sum-rate versus number of subcarriers.}
\end{figure}

Finally, Fig. 11 shows the change of system sum-rate under different algorithms when the number of RIS transmissive elements increases. We can see that the performance of the proposed algorithm still outperforms other benchmark algorithms. Moreover, the system sum-rate increases as the number of RIS transmissive elements increases. This is because the RIS transmissive element will strengthen the user’s uplink transmission signal and bring a certain gain, which is positively correlated with the number of RIS transmissive elements. Therefore, the more RIS transmissive elements, the more gain it brings, and the greater the system sum-rate.
It provides a good guide for multi-antenna systems based on transmissive RIS, which can improve system performance by increasing the number of low-cost transmissive elements, which has great potential for future 6G networks that needs to reduce costs and power consumption.
\begin{figure}
	\centerline{\includegraphics[width=8cm]{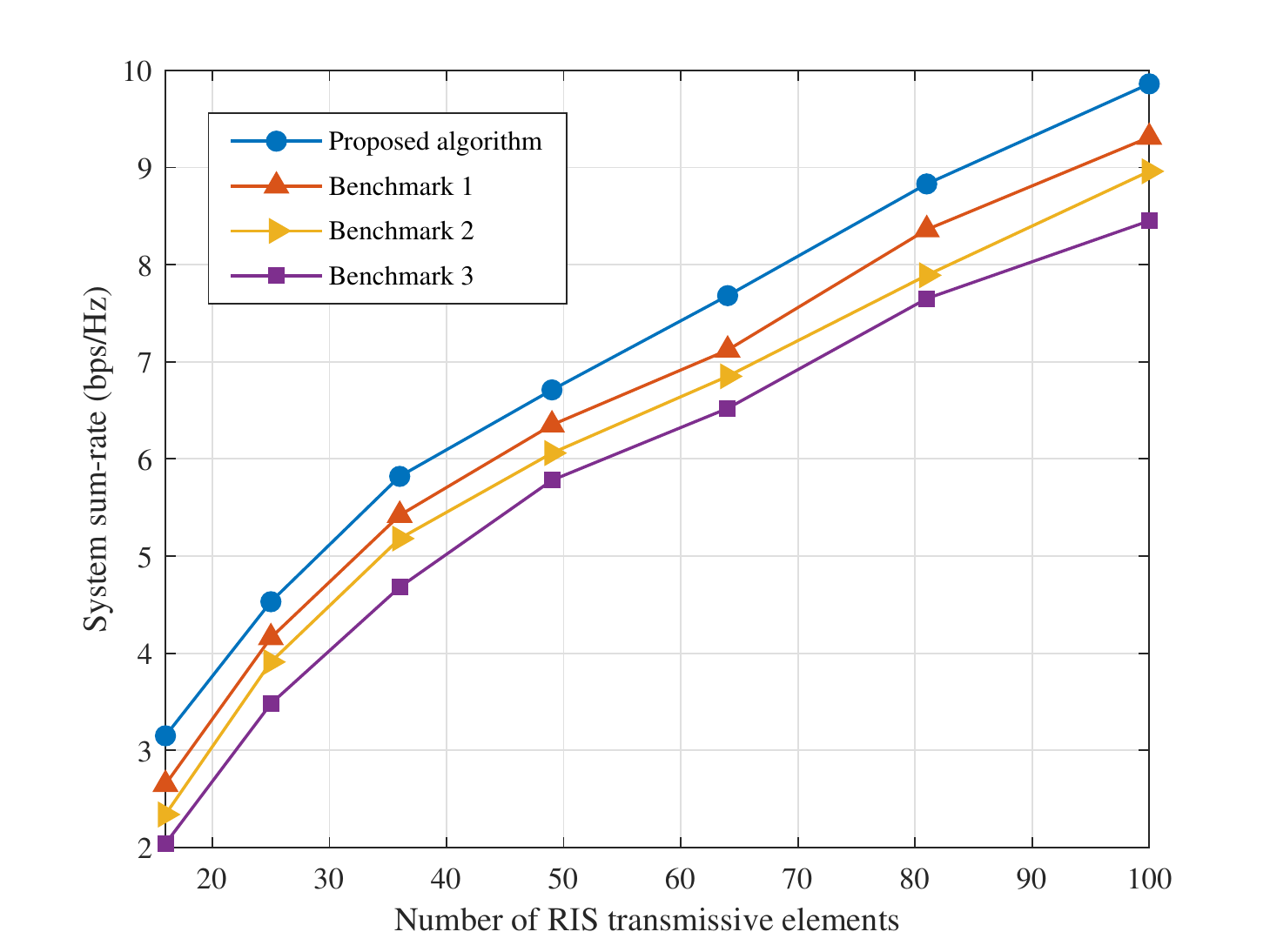}}
	\caption{System sum-rate versus number of RIS transmissive element.}
\end{figure}

\section{Conclusions}
The design and optimization in the transmissive RIS transceiver enabled uplink communication system via OFDMA has been investigated in this paper. First, we propose a novel transmissive RIS-based transceiver architecture and illustrate its advantages due to the structure. Then, for the proposed system, a problem of maximizing system sum-rate by jointly optimizing subcarrier allocation, power allocation, and RIS transmissive coefficient design is formulated. In order to solve this challenging problem, the original problem is decomposed into two sub-problems by the AO algorithm and solved sequentially. First, for the transformed problem, a joint optimization of multi-user power allocation and subcarrier allocation algorithm based on Lagrangian dual decomposition is proposed. Then, based on the obtained power allocation and subcarrier allocation schemes, the RIS transmissive coefficient is obtained via DC programming and penalty function methods. Finally, the two sub-problems are iterated alternately until convergence is achieved. Next, the convergence and computational complexity are also given. Fianlly, numerical simulation results verify that our proposed algorithm can improve system sum-rate, and can improve system performance by increasing the number of RIS transmissive elements, which is promising for reducing power consumption and cost in future networks.

\ifCLASSOPTIONcaptionsoff
  \newpage
\fi



\bibliographystyle{IEEEtran}
\bibliography{reference}
\end{document}